\documentclass[prl,superscriptaddress,amsmath,amssymb,twocolumn,aps]{revtex4-2}
\usepackage{natbib}
\usepackage{graphicx}         
\usepackage{dcolumn}          
\usepackage{bm}              
\usepackage{upgreek}
\usepackage{booktabs} 				
\usepackage{tabularx}
\usepackage{array}
\usepackage{ulem}
\usepackage{float}
\usepackage[colorlinks=true,urlcolor=blue,breaklinks=true]{hyperref}
\usepackage{xcolor}
\usepackage{subfigure}
\usepackage{orcidlink}
\usepackage[a4paper,left=2cm,right=2cm]{geometry}

\usepackage{float}
\usepackage[none]{hyphenat}
\usepackage{siunitx}
\usepackage{stackengine}
\usepackage{placeins}

\usepackage{ragged2e}

\usepackage{titlesec}

\begin{document}
\title{Frustrated Frustration of Arrays with Four-Terminal Nb-Pt-Nb Josephson Junctions}

\author{Justus Teller\;\orcidlink{0000-0003-0032-4200}}
\email{j.teller@fz-juelich.de}
\affiliation{Peter Gr\"unberg Institut (PGI-9), Forschungszentrum J\"ulich, 52425 J\"ulich, Germany}
\affiliation{JARA-Fundamentals of Future Information Technology, J\"ulich-Aachen Research Alliance, Forschungszentrum J\"ulich and RWTH Aachen University, Germany}

\author{Christian Schäfer\;\orcidlink{0009-0007-0692-1205}}
\affiliation{Peter Gr\"unberg Institut (PGI-9), Forschungszentrum J\"ulich, 52425 J\"ulich, Germany}
\affiliation{JARA-Fundamentals of Future Information Technology, J\"ulich-Aachen Research Alliance, Forschungszentrum J\"ulich and RWTH Aachen University, Germany}

\author{Kristof Moors\;\orcidlink{0000-0002-8682-5286}}
    \thanks{Present Address: Imec, Kapeldreef 75, 3001 Leuven, Belgium}
\affiliation{Peter Gr\"unberg Institut (PGI-9), Forschungszentrum J\"ulich, 52425 J\"ulich, Germany}
\affiliation{JARA-Fundamentals of Future Information Technology, J\"ulich-Aachen Research Alliance, Forschungszentrum J\"ulich and RWTH Aachen University, Germany}

\author{Benjamin Bennemann}
\affiliation{Peter Gr\"unberg Institut (PGI-10), Forschungszentrum J\"ulich, 52425 J\"ulich, Germany}

\author{Matvey Lyatti\;\orcidlink{0000-0003-0837-4886}}
\affiliation{Peter Gr\"unberg Institut (PGI-9), Forschungszentrum J\"ulich, 52425 J\"ulich, Germany}
\affiliation{JARA-Fundamentals of Future Information Technology, J\"ulich-Aachen Research Alliance, Forschungszentrum J\"ulich and RWTH Aachen University, Germany}

\author{Florian Lentz\;\orcidlink{0000-0002-8716-6446}}
\affiliation{Helmholtz Nano Facility, Forschungszentrum J\"ulich, J\"ulich 52425, Germany}

\author{Detlev Gr\"utzmacher\;\orcidlink{0000-0001-6290-9672}}
\affiliation{Peter Gr\"unberg Institut (PGI-9), Forschungszentrum J\"ulich, 52425 J\"ulich, Germany}
\affiliation{JARA-Fundamentals of Future Information Technology, J\"ulich-Aachen Research Alliance, Forschungszentrum J\"ulich and RWTH Aachen University, Germany}

\author{Roman-Pascal Riwar\;\orcidlink{0000-0002-0402-1090}}
\affiliation{Peter Gr\"unberg Institut (PGI-2), Forschungszentrum J\"ulich, 52425 J\"ulich, Germany}

\author{Thomas Sch\"apers\;\orcidlink{0000-0001-7861-5003}}
\email{th.schaepers@fz-juelich.de}
\affiliation{Peter Gr\"unberg Institut (PGI-9), Forschungszentrum J\"ulich, 52425 J\"ulich, Germany}
\affiliation{JARA-Fundamentals of Future Information Technology, J\"ulich-Aachen Research Alliance, Forschungszentrum J\"ulich and RWTH Aachen University, Germany}
\hyphenation{}
\date{\today}

\begin{abstract}
We study the frustration pattern of a square lattice with in-situ fabricated Nb-Pt-Nb four-terminal Josephson junctions. The four-terminal geometry gives rise to a checker board pattern of alternating fluxes $f$, $f^\prime$ piercing the plaquettes, which stabilizes the Berezinskii--Kosterlitz--Thouless transition even at irrational flux quanta per plaquette, due to an unequal repartition of integer flux sum $f+f^\prime$ into alternating plaquettes. This type of frustrated frustration manifests as a beating pattern of the dc resistance, with state configurations at the resistance dips gradually changing between the conventional zero-flux and half-flux states. Hence, the four-terminal Josephson junction array offers a promising platform to study previously unexplored flux and vortex configurations, and provides an estimate on the spatial expansion of the four-terminal Josephson junction central weak link area.

\end{abstract}
\maketitle

\textit{Introduction}. Arrays of Josephson junctions have been studied since the 1980s~\cite{h.sanchez_properties_1981} and a broad base of knowledge about the physics of these arrays has been accumulated over the years~\cite{abraham_resistive_1982a, lobb_theoretical_1983a, teitel_josephsonjunction_1983, benz_fractional_1990, rzchowski_vortex_1990, mooij_unbinding_1990, fazio_charge_1991, geigenmuller_friction_1993, dang_vortex_1993, martinoli_two_2000, newrock_twodimensional_2000, fazio_quantum_2001, tinkhambook, bruder_bosehubbard_2005, romito_solidstate_2005, gladchenko_superconducting_2009, terhal_majorana_2012, houck_onchip_2012}. Recent findings in the field include the engineering of energy-phase relations with arrays~\cite{bozkurt_doublefourier_2023}, a deeper understanding of the vortex-lattice states in arrays~\cite{penner_resistivity_2023}, the demonstration of giant fractional Shapiro steps in anisotropic arrays~\cite{panghotra_giant_2020}, and the creation of arrays made of superconducting islands on a normal-conducting weak link material~\cite{panghotra_giant_2020, song_interference_2023, bottcher_dynamical_2023, vervoort_dcoperated_2024, reinhardt_spontaneous_2025}.

Typically, Josephson junction arrays are formed by two-terminal junctions. Recently, however, multi-terminal Josephson junctions received increasing attention~\cite{pfeffer_subgap_2014, riwar_multiterminal_2016, draelos_supercurrent_2019a, graziano_transport_2020, pankratova_multiterminal_2020}. In general, a multi-terminal Josephson junction is defined by multiple superconducting leads being connected by a central weak link region~\cite{pankratova_multiterminal_2020, riwar_multiterminal_2016}. Various weak-link materials can be used for these multi-terminal Josephson junctions. 

There is a broad spectrum of studies of devices with weak links such as semiconductors~\cite{graziano_selective_2022, coraiola_phaseengineering_2023, gupta_gatetunable_2023, coraiola_fluxtunable_2024, gupta_evidence_2024}, graphene~\cite{draelos_supercurrent_2019a, arnault_multiterminal_2021a, chiles_nonreciprocal_2023, arnault_dynamical_2022}, and topological insulators~\cite{kolzer_supercurrent_2023, behner_superconductive_2025, kudriashov_nonreciprocal_2025, nikodem_large_2025}.
Topological insulator based multi-terminal Josephson junctions are considered as key element in various Majorana fermion braiding architectures~\cite{fu_superconducting_2008, vanheck_coulombassisted_2012, fulga_effects_2013}. More generally, a multi-terminal Josephson junction is predicted to host topological states without requiring any topological material~\cite{riwar_multiterminal_2016}. 

\begin{figure}[hbtp]
\includegraphics[width=0.48\textwidth]{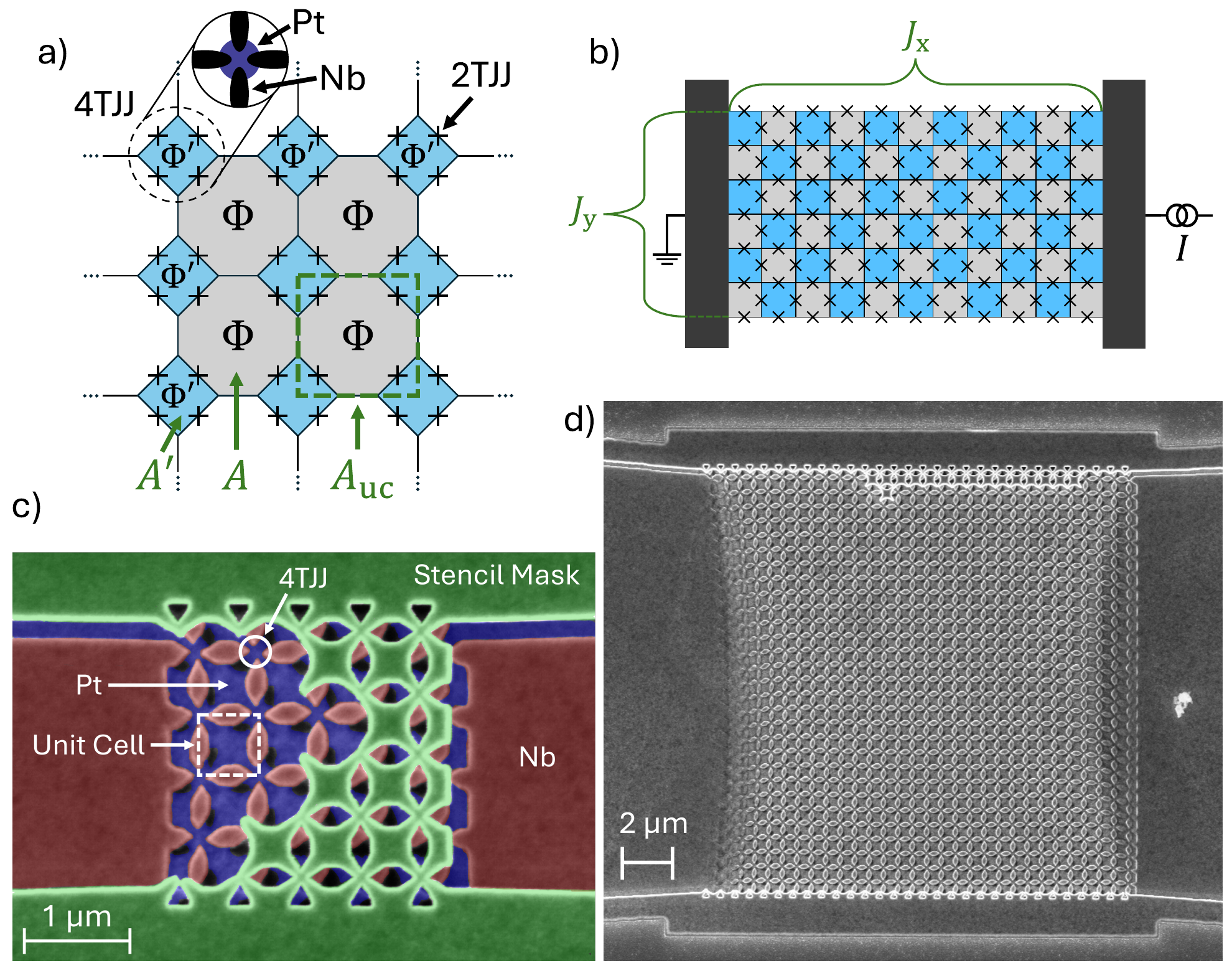}
\caption{a) Schematics of the Josephson junction array where each 4TJJ is described by four interconnected two-terminal Josephson junctions (see dashed circle)~\cite{draelos_supercurrent_2019a, graziano_transport_2020, arnault_multiterminal_2021a, arnault_dynamical_2022}. Their weak link material is platinum (see inset). The array is formed by connecting the superconducting arms of the 4TJJs. Upon application of a magnetic field each large plaquette area ($A$) of the array is penetrated by a magnetic flux $\Phi$ and the 4TJJ weak link area ($A^\prime$) by a flux $\Phi^\prime$. b) Theoretical representation of the array shown in a) as a 2TJJ array with square plaquettes having alternating frustrations $f = \Phi/ \Phi_0$ (grey) and $f^\prime = \Phi^\prime/ \Phi_0$ (blue). c) False-color scanning electron micrograph of a $5\times5$ four-terminal Josephson junction array with partially removed stencil mask. The platinum (blue) is deposited under rotation and, thus, covers a larger area. The niobium (red) forms the superconducting contacts. The dashed square represents unit cell area $A_\text{uc}$. d) Scanning electron micrograph of the $30\times30$ 4TJJ array presented in this study. Due to the shadow evaporation process, the 4TJJ array is slightly deformed at the left and right ends of the array.\label{FIG:INT_Schematic}}
\end{figure}

We here report on the fabrication and study of a square array comprised of \textit{multi-terminal} Josephson junctions, as illustrated in Fig.~\ref{FIG:INT_Schematic}~a).
In contrast to conventional Josephson junction arrays, where a two-terminal Josephson junction (2TJJ) is placed in each arm forming a plaquette of the array, here, a four-terminal Josephson junction (4TJJ) is located in each corner of the unit cell. As it turns out, this system has not a single flux parameter (as usual arrays do) but an alternating pattern of two fluxes. This allows us to find a stable superconducting phase for the array, even for irrational fluxes piercing the individual plaquettes. Our work thus connects vortex dynamics and the Berezinskii--Kosterlitz--Thouless (BKT) transition to the emerging field of incommensurable, quasiperiodic physics in solid state systems~\cite{Ahn2018,FuPixley2020,andrei2021marvels,mak2022semiconductor, UriJarillo2023,Herrig_2025}. While the usual notion of quasiperiodic materials refers to lattices in real space, the incommensurability here is detectable in the space spanned by the applied flux.

\textit{Experiments}. Our 4TJJ array is based on a superconductor/normal conductor structure, using a metal as a weak link between the superconducting electrodes \cite{h.sanchez_properties_1981, rzchowski_vortex_1990, pfeffer_subgap_2014, panghotra_giant_2020, skryabina_josephson_2017, vervoort_dcoperated_2024}. More specifically, we use Nb for the four closely spaced superconducting electrodes connected by a metallic Pt weak link. The manufacturing process of the 4TJJ array is based on a stencil lithography process using molecular beam epitaxy with a high device yield by ensuring ultra-clean interfaces between the Nb and Pt layers \cite{schuffelgen_selective_2019} (see Fig.~\ref{FIG:INT_Schematic} and \cite{supplementary}).\\

 Figure~\ref{FIG:INT_Schematic} d) shows a scanning electron microscope image of the $30\times30$ 4TJJ array investigated in this study. In addition to the device presented in the main text, we measured an identical array device and a reference two-terminal Josephson junction, all fabricated on the same substrate during the same fabrication run~\cite{supplementary}.\\

Magnetotransport measurements have been performed on a $30\times30$ 4TJJ array at a temperature of 80\:mK. Without external magnetic field, the array shows a critical current of $I_\mathrm{c} = 57\;\mu$A and, close to the superconducting regime, the device has a differential resistance of around 5.5\;$\Omega$~\cite{supplementary}. 

\begin{figure}[hbtp!]
\centering
\includegraphics[width=0.45\textwidth]{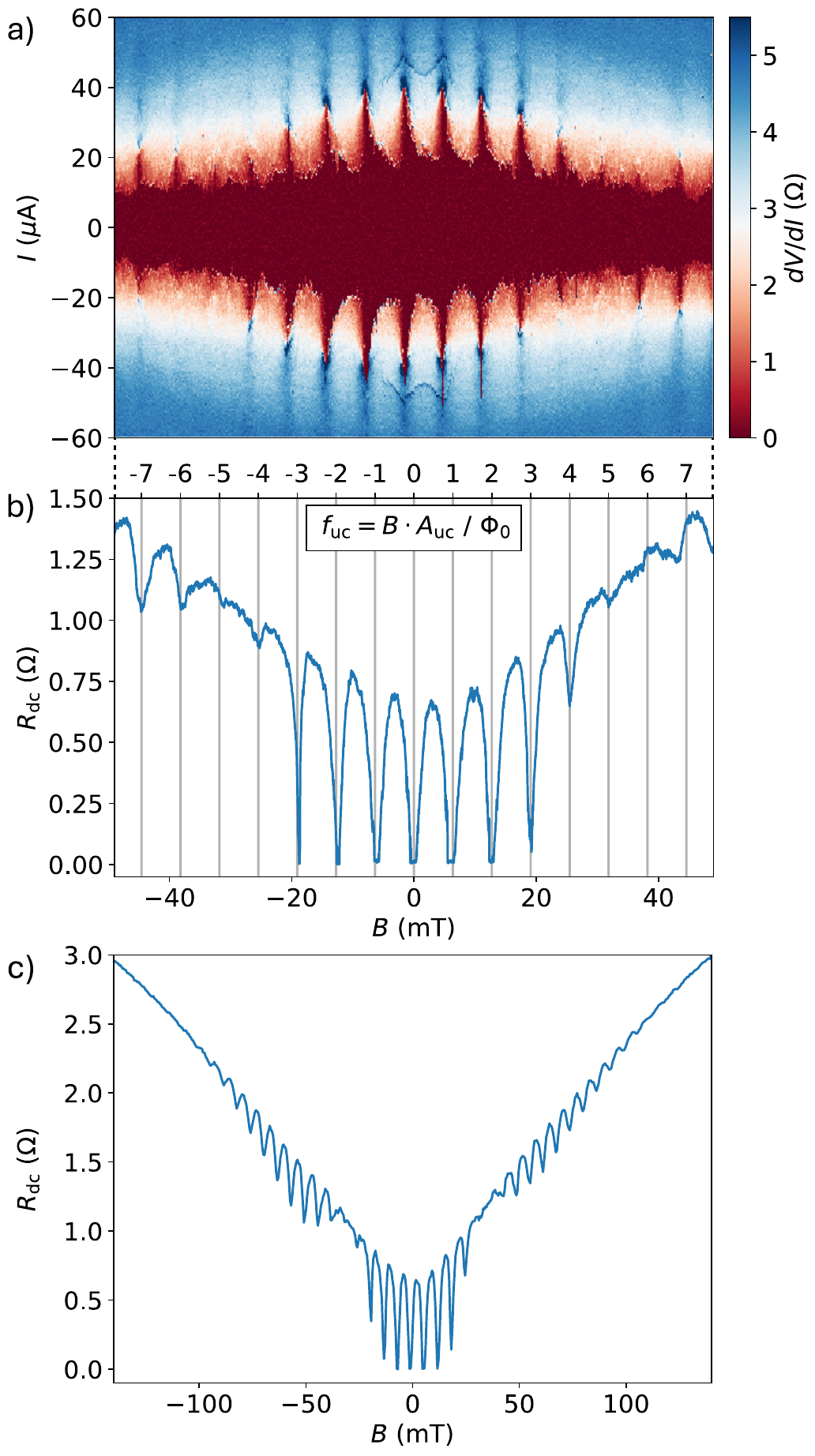}
\caption{Measurement data of the $30\times30$ 4TJJ array at 80\;mK. a) Differential resistance as a function of bias current and magnetic field. A periodic oscillation of the critical current is clearly visible. b) Resistance as a function of magnetic field with an applied dc bias of 30\;$\mu$A. The resistance oscillations correspond to a periodicity of 6.25\;mT. At around $\pm5\;f_\mathrm{uc}$, the resistance oscillations are damped. c) Resistance oscillations under magnetic field ranging to $\pm140$\;mT with an applied dc bias of 30\;$\mu$A. In total, 30 flux quantum oscillations are present.\label{FIG:Array_transport_data}}
\end{figure}

The differential resistance vs. bias current and out-of-plane magnetic field depicted in Fig.~\ref{FIG:Array_transport_data}~a) shows periodic oscillations of the critical current with magnetic field. When applying a fixed dc current of 30\;$\mu$A through the device, its resistance oscillates with the same periodicity in magnetic field [cf.~Fig.~\ref{FIG:Array_transport_data}~b)], which was determined to be 6.25\;mT by a fast Fourier transform~\cite{supplementary}. In addition, a device- and array-independent magnetic hysteresis of the resistance pattern was measured, also present in the reference 2TJJ and both arrays, which is not discussed further in the main text (see Sec.~V in the supplementary material).

To describe the properties of Josephson junction arrays, the so-called frustration parameter $f = B\cdot A/\Phi_0$, with $A$ being the respective plaquette area, $B$ the magnetic field strength perpendicular to the array plane, and $\Phi_0 = h/2e$ the magnetic flux quantum, is used to characterize the magnetic resistance pattern, i.e., frustration pattern, see e.g. Refs.~\cite{rzchowski_vortex_1990, song_interference_2023}. It describes the average number of flux quanta piercing through an array plaquette. For rescaling the magnetic field into frustration, the unit cell area, as indicated in Fig.~\ref{FIG:INT_Schematic}~c), has been determined to be $A_\mathrm{uc} = (570\;$nm$)^2$ by scanning electron microscopy~\cite{supplementary}. This unit cell is the sum of a large plaquette area $A$ (with $\Phi = B\cdot A$) and a small plaquette area $A'$ (with $\Phi' = B\cdot A'$) of Fig.~\ref{FIG:INT_Schematic}~a), i.e., $A_\text{uc} = A + A'$ (with $\Phi_\text{uc} = B\cdot A_\text{uc}$). The flux quantum oscillations of the resistance fit to $A_\text{uc}$ in Fig.~\ref{FIG:Array_transport_data} b).

The resistance signal is similar to what, e.g. \mbox{Rzchowski}~\textit{et al.}~\cite{rzchowski_vortex_1990} measured, however, the expected frustration pattern for rational values of $f$ is missing. As can be seen in Fig.~\ref{FIG:Array_transport_data} c), at certain magnetic field around $\pm5$\;$f_\mathrm{uc}$, the oscillations first disappear and then reappear when further increasing the magnetic field. The resistance oscillations can be seen to magnetic fields above $\pm100$\;mT with, in total, 30 flux quantum oscillations.\\

\textit{Theoretical discussion}. To explain the measured behavior of the 4TJJ array, we introduce a corresponding theoretical model. Based on the resistively capacitively shunted junction (RCSJ) network model for multi-terminal Josephson junctions~\cite{draelos_supercurrent_2019a, graziano_transport_2020, arnault_multiterminal_2021a, arnault_dynamical_2022}, the array can be described, as shown in Fig.~\ref{FIG:INT_Schematic}~a), with four 2TJJs creating one 4TJJ~\cite{supplementary}. This introduces a second lattice of areas pierced by magnetic flux, the central weak link region of the 4TJJs.
The multi-terminal array model can therefore be conveniently represented in terms of an ordinary $J_x\times J_y$ square lattice junction model [where $J_x$ and $J_y$ are the horizontal and vertical number of superconducting nodes, see Fig.~\ref{FIG:INT_Schematic}~b)], with the following important difference to previous theoretical and experimental studies: instead of all the plaquettes being pierced by the same flux, we get an alternating (checker board) flux pattern ($f=\Phi/\Phi_0$ and $f^\prime=\Phi^\prime/\Phi_0$). We deploy a classical RCSJ model approach, where the equations of motion for the superconducting phases at all nodes follow as usual from the Kirchhoff laws~\cite{supplementary}. We focus on the overdamped regime~\footnote{Using the extracted values for individual junction capacitances, resistances, and critical currents, we estimate the Steward-McCumber parameter to be $\approx 5.4\times 10^{-6}$, well within the overdamped regime.}, neglecting the capacitive contributions to the equations of motion, and assume zero temperature. As will become apparent below, it is instructive to include a finite Fraunhofer effect~\cite{tinkhambook} for the individual junctions in the lattice.

\begin{figure}
    \centering
    \includegraphics[width=0.99\linewidth]{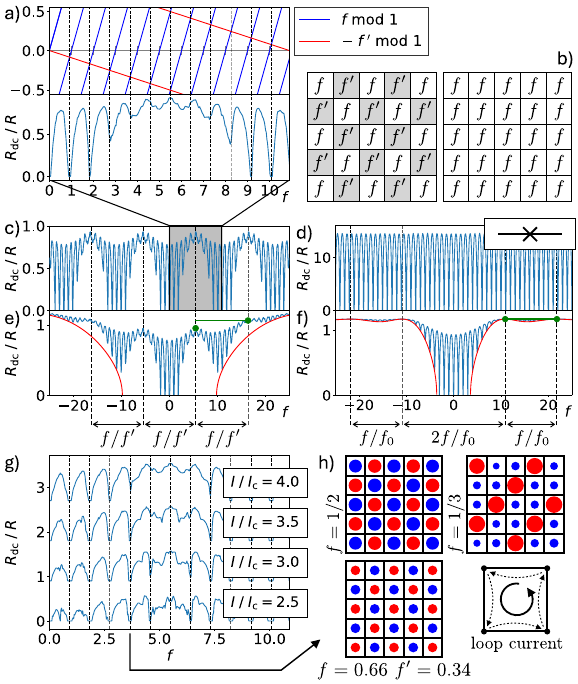}
    \caption{Results of the theoretical analysis. In all curves (a,c-g), the $x$-axis is $f$, and the $y$-axis is $R_\text{dc}/R$. (a) $R_\text{dc}$ as a function of $f$ for the checker board model with $f/f^\prime=10.9$ and $I/I_c=5.0$. (b) Checker board ($f\neq f^\prime$) versus regular ($f=f^\prime$) lattice models. (c) Zoomed out version of (a) showing the beating pattern. (d) $R_\text{dc}$ for the regular lattice model $f=f^\prime$, with all other parameters the same as in (a,c). (e,f) $R_\text{dc}$ including Fraunhofer effect (red curves, see main text), where for the checker board model $f_0=30$ (e) and for the regular lattice $f_0=10.9$ (f). (g) $R_\text{dc}$ for $f/f^\prime=10.9$ (no Fraunhofer) for decreasing bias current. Data shifted for clarification. (h) Equilibrium loop current configuration ($I=0$) for different values of $f,f^\prime$. The dots indicate counter- (red) or clockwise (blue) going currents and their size represents the magnitude of the loop current (relative linear scale).}
    \label{fig:theory}
\end{figure}

Including a source and drain contact on two sides of the lattice (across which the bias current $I$ is applied), the RCSJ model allows for a direct computation of the dc resistance $R_\text{dc}$ as a function of the magnetic field. We do so by an explicit numerical evaluation of the classical RCSJ equations of motion~\footnote{We use the NDSolve routine on Mathematica for explicit evaluation of the equations of mtion.}. The results of the simulation are summarized in Fig.~\ref{fig:theory}. For computational simplicity, the calculations were performed on a small array of $J_x=J_y=6$.

It is instructive to first consider the special case $f,f'\in\mathbb{Z}$ (equivalent to $f=f^\prime=0$). Here, the dc resistance can be found analytically (as both quantum and thermal phase slips are absent)~\cite{supplementary},
\begin{equation}\label{eq_Rdc_null}
    \frac{R_\text{dc}}{R}=(J_x+1)\sqrt{\frac{1}{J_y^2}-\frac{I_c^2}{I^2}}\ ,
\end{equation}
where $R$ and $I_c$ are the individual junction resistance and critical current, respectively. The array thus transitions from superconducting ($R_\text{dc}=0$) to resistive ($R_\text{dc}>0$) when the bias current exceeds $J_yI_c$. For a regular square lattice [$f=f^\prime$, see also Fig.~\ref{fig:theory}~b)] it is  well known that the array generally leaves the superconducting regime for noninteger values of $f$, even though $I<J_y I_c$. For sufficiently low temperatures and current biases, the dc resistance nonetheless experiences dips at special rational values -- known as the frustration pattern.

Consider now the general checker board array, $f\neq f^\prime$ [left panel in Fig.~\ref{fig:theory}~b)], where we denote the (constant) ratio $\beta=f/f^\prime>1$. Here, there emerges a beating pattern, with the two characteristic periods $\beta>1$ and $\beta/(1+\beta)<1$ [see Figs.~\ref{fig:theory}~a)~and~c)]. The larger period ($\beta$) is in general only approximate; e.g., for incommensurate $\beta$ the system can only approximately return to mutually integer $f,f^\prime$ -- exhibiting a quasiperiodicity in flux space. The irrational nature of $\beta$ is however also highly relevant for the smaller period (which is exact): it separates the points where the sum of two neighbouring plaquette fluxes is zero, i.e., $f+f^\prime\in\mathbb{Z}$ [see top inset in Fig.~\ref{fig:theory}~a)]. For the experimentally extracted value of $\beta\approx 10.9$ (cf. Fig.~\ref{FIG:Array_transport_data},~\cite{supplementary}) the difference between the two frequencies is large, leading to a clearly visible beating pattern [Fig.~\ref{fig:theory}~c)]. Once $\beta$ is known, the central weak link area of a 4TJJ can be determined via $A' = A_\text{uc}/(1+\beta)$, leading to $A' \approx (165\;\text{nm})^2$~\cite{supplementary}. Crucially, for integer $f+f^\prime$, the repartition of the total flux into the two neighbouring plaquettes ($f$ and $f^\prime$) does not need to occur for special (integer or rational) value, since the ratio of the two fluxes, $\beta$, is in general incommesurate. Consequently, the system exhibits a stabilization of superconducting (BKT) phase (dips in $R_\text{dc}$) even for \textit{irrational} $f,f^\prime$, a feature we choose to name 'frustrated frustration'. 

We now include the Fraunhofer effect, $I_c\sim \text{sinc}(\pi f/f_0)$, where the parameter $f_0$ captures the junction area. The resulting reduction of the critical current at finite $f$ leads to a base offset in the $R_\text{dc}$-curve, see Fig.~\ref{fig:theory}~e) [the red curve represents Eq.~\eqref{eq_Rdc_null} with $f$-dependent $I_c$], in remarkable resemblance to the experimental data, Fig.~\ref{FIG:Array_transport_data}~c). As a comparison, take the regular square lattice, $f=f^\prime$, where a qualitatively very similar beating pattern arises due to Fraunhofer [Fig.~\ref{fig:theory}~d) versus Fig.~\ref{fig:theory}~f)], by setting $f_0$ (instead of $f/f^\prime$) to $\approx 10.9$. However, this alternative model can be safely excluded to explain the experimentally observed phenomenology. First, the Fraunhofer beating pattern in Fig.~\ref{fig:theory}~f) skips a beat at $f=0$, due to the $\text{sinc}$-function having no zero at the origin, such that there are two beating pattern frequencies, $2/f_0$ (main beat around $f=0$) and $1/f_0$ (side beats) [see Fig.~\ref{fig:theory}~f)]. Figure~\ref{FIG:Array_transport_data} only exhibits one beating frequency, in alignment with the checker board model~\ref{fig:theory}~c). Moreover, the area ratio extracted from micrographs of the device align much better with the first model ($f/f^\prime\approx 10.9$ and much larger $\sim f_0$). Related to that, for $f/f_0\ll 1$ (while $I\gg JI_c$), the base offset is approximately linear, $R_\text{dc}\sim f$, consistent with the measured asymptotics in Fig.~\ref{FIG:Array_transport_data}. Indeed, nodes separating two beats [green dots in Figs.~\ref{fig:theory}~e)~and~f)] can only exhibit an increase in $R_\text{dc}$ in the checker board model, while in the regular lattice, the onset of the beating pattern must coincide with $R_\text{dc}$ reaching the constant plateau value $\approx R$.

We observe that the pattern in Fig.~\ref{fig:theory}~a) transitions from resistance dips to peaks at integer $f+f^\prime$, depending on whether $f$ and $f^\prime$ are individually closer to integer or half-integer. Those peaks can become dips when lowering the bias current, see Fig.~\ref{fig:theory}~g), marking the onset of the checker board version of regular frustration -- which goes hand in hand with the formation of additional resistance dips at $f+f^\prime$ \textit{half} integer [see, $0<f<1$ in Fig.~\ref{fig:theory}~g)]~\footnote{Such extra features have not been detected in experiment, likely due to application of higher bias currents. Future experiments could be dedicated to the search of additional frustration features.}. Overall, the 'frustrated frustration' pattern ($f+f^\prime \in \mathbb{Z}$) is more stable than regular frustration. For an intuitive understanding, consider the equilibrium ($I=0$) configurations of loop currents~\footnote{Our definition of loop currents, as shown in Fig.~\ref{fig:theory}~h), is related to, but distinct from, the more commonly considered vortex picture.} flowing through single plaquettes, see Fig.~\ref{fig:theory}~h). Thus we find that, e.g., the resistance dip at $f\approx 2/3,f^\prime \approx 1/3$ resembles much more closely the configuration of $f=f^\prime=1/2$ than the one at $f=f^\prime=1/3$ -- however, with an overall reduced loop current magnitude, indicating increased stability. Stability considerations of the various frustration features can also be understood in terms of the regular vortex model~\cite{supplementary}.

\textit{Conclusion}. We demonstrated an array made with 4TJJs. For this we introduced an in-situ fabrication technique for arrays. A frustration pattern was measured which differs from the expectation for ordinary 2TJJ arrays. The difference could be explained by theoretically introducing a checker board lattice with two alternating flux patterns $f$ and $f^\prime$. The periodicity of the beating pattern can be connected to the area ratios $A$/$A'$. In particular, while dips in the dc resistance can be linked to the sum $f+f^\prime$ being integer, the in general irrational area ratio leads to an irrational repartition of this total flux into the alternating plaquettes. This allows for the stabilization of the superconducting BKT phase at irrational fractions of flux penetrating individual plaquettes. This feature allows us to estimate the spatial expansion of the 4TJJ central weak link region, the result of which is compatible with the junction geometry. Overall, the here considered setup opens up a previously unexplored class of array system with alternating flux textures. Given that detailed understanding of vortex lattice structures and their dynamics are still an open question for regular lattices with a single flux $f$~\cite{penner_resistivity_2023},  the here introduced alternating $f, f^\prime$ lattice will likely open up a new research direction in its own right (e.g., exploring the impact of capacitive quantum fluctuations on the phase diagram as in Ref.~\cite{fazio_charge_1991}, now including incommensurate vortex features). 

\par
\textit{Acknowledgements}. We thank Herbert Kertz for technical assistance. All samples have been prepared at the Helmholtz Nano Facility~\cite{albrecht_hnf_2017}. This work is funded by the Deutsche Forschungsgemeinschaft (DFG, German Research Foundation) under Germany's Excellence Strategy – Cluster of Excellence Matter and Light for Quantum Computing (ML4Q) EXC 2004/1 – 390534769, by the German Federal Ministry of Education and Research (BMBF) via the Quantum Futur project ‘MajoranaChips’ (Grant No. 13N15264) and the Quantum Futur project with grant no. 13N14891 (RR), as well as the Bavarian Ministry of Economic Affairs, Regional Development and Energy within Bavaria’s High-Tech Agenda Project ”Bausteine für das Quantencomputing auf Basis topologischer Materialien mit experimentellen und theoretischen Ansätzen” (grant no. 07 02/686 58/1/21 1/22 2/23).

\let\oldaddcontentsline\addcontentsline
\renewcommand{\addcontentsline}[3]{}

\let\addcontentsline\oldaddcontentsline

\clearpage
\widetext

\titleformat{\section}[hang]{\bfseries}{\MakeUppercase{Supplemental Note} \thesection:\ }{0pt}{\MakeUppercase}

\setcounter{section}{0}
\setcounter{equation}{0}
\setcounter{figure}{0}
\setcounter{table}{0}
\setcounter{page}{1}
\renewcommand{\thesection}{\arabic{section}}
\renewcommand{\thesubsection}{\Alph{subsection}}
\renewcommand{\theequation}{S\arabic{equation}}
\renewcommand{\thefigure}{S\arabic{figure}}
\renewcommand{\figurename}{Supplemental Figure}
\renewcommand{\tablename}{Supplemental Table}
\renewcommand{\bibnumfmt}[1]{[S#1]}
\renewcommand{\citenumfont}[1]{S#1}

\begin{center}
	\textbf{\large Frustrated Frustration of Arrays with Four-Terminal Nb-Pt-Nb Josephson Junctions\\(Supplementary Material)}
\end{center}

{
	\hypersetup{linkcolor=black}
	\tableofcontents
}

\section{Fabrication procedure}
The stencil lithography fabrication process of the four-terminal
Josephson junction (4TJJ) array with molecular beam epitaxy is described in the following. As a first step, a Si(111) substrate is covered by a 300-nm-thick SiO$_2$ layer followed by a 100-nm-thick Si$_3$N$_4$ layer. Subsequently, the stencil mask is patterned into the Si$_3$N$_4$ layer by electron beam lithography and reactive ion etching. As a next step, the stencil mask is under-etched by hydrofluoric acid, removing the SiO$_2$ layer below and revealing the Si(111) substrate surface. The stencil mask can be seen in main text Fig.~\ref{FIG:INT_Schematic}~c). After inserting the pre-patterned substrate into an ultra high vacuum growth chamber, a 10-nm-thick Al$_2$O$_3$ interlayer is first deposited under rotation. The rotation ensures that the junction array area is completely covered by Al$_2$O$_3$. Next, a 20-nm-thick Pt layer is deposited under rotation. Here, the deposition at different angles due to the substrate rotation results in the deposition of Pt connecting the openings of the stencil mask. Finally, a 50-nm-thick superconducting Nb layer is deposited without rotation, resulting in a Nb electrode pattern defined by the opening of the stencil mask. As a last step, a 5-nm-thick Al$_2$O$_3$ capping layer is deposited under rotation. In main text Fig.~\ref{FIG:INT_Schematic}~c), a scanning  electron microscope image of the final device structure is shown. The four-terminal Josephson junction consists of four superconducting Nb islands connected at their tips by a central Pt weak link area. After the material deposition, the stencil mask is removed by applying ultrasonic power. The square unit cell of the array formed by a 4TJJ in each corner is indicated in main text Fig.~\ref{FIG:INT_Schematic}~c) as well. To prevent screening of magnetic fluxes by screening currents in superconducting material surrounding the array, the Nb layer outside the device is removed by reactive ion etching (see suppl. section~\ref{sec:SUPP_Array_Device_Description}).  

\section{Measurement setup}
The measurements have been performed in a dilution refrigerator with a base temperature of 70\;mK equipped with a superconducting magnet. For measurements with magnetic fields below 80\;mT, an external high-resolution current source was used to supply the bias current of the magnet. The voltage signal, from which the resistance ($R_\text{dc}$) and differential resistance ($dV/dI$) has been calculated, was measured with a quasi 4-point setup by applying a dc bias to the array. In additional measurements (suppl. section \ref{sec:SUPP_Misc_Measurements}), the dc signal was superimposed with a smaller AC signal in order to measure the differential resistance using the lock-in technique and the ac signal has been used alone for ac resistance measurements dependent on magnetic field.

\section{Second 30$\times$30 4TJJ array}
\begin{figure}[h]
	\centering
	\subfigure[]{\includegraphics[width=0.55\textwidth]{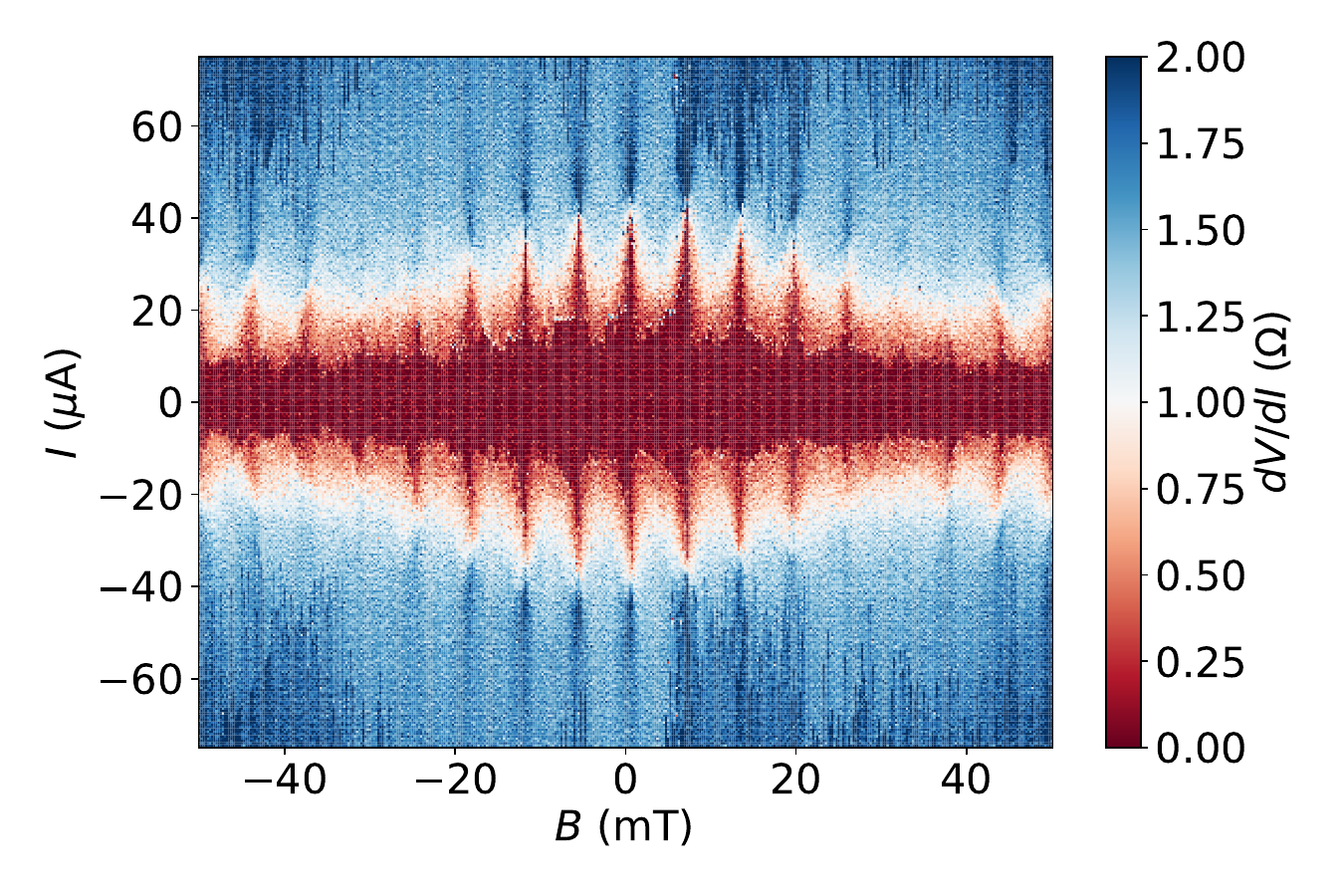}}\quad
	\subfigure[]{\includegraphics[width=0.38\textwidth]{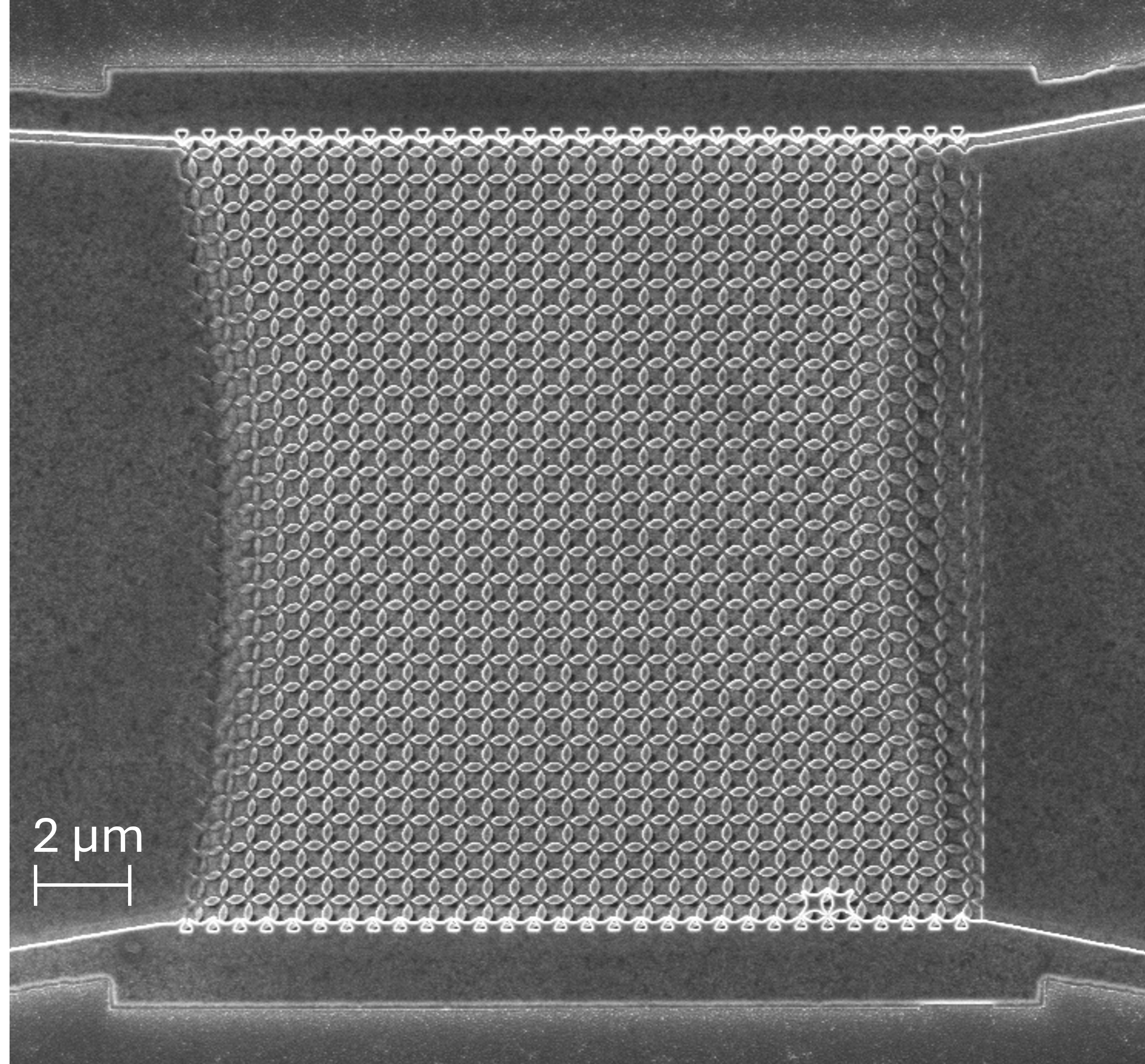}}\\
	\subfigure[]{\includegraphics[width=0.48\textwidth]{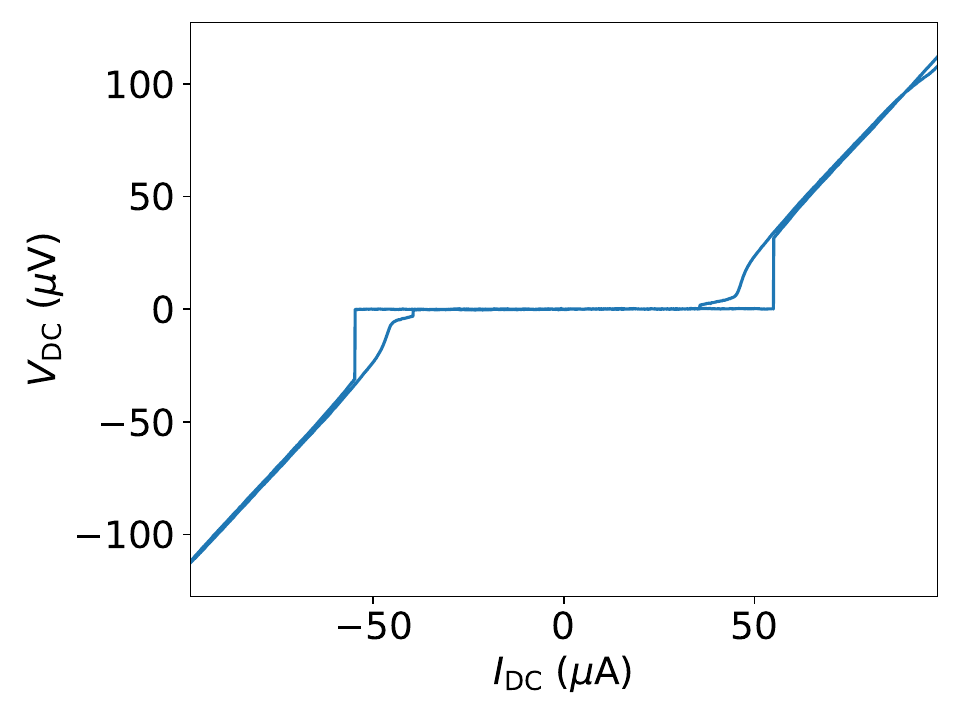}}\quad
	\subfigure[]{\includegraphics[width=0.48\textwidth]{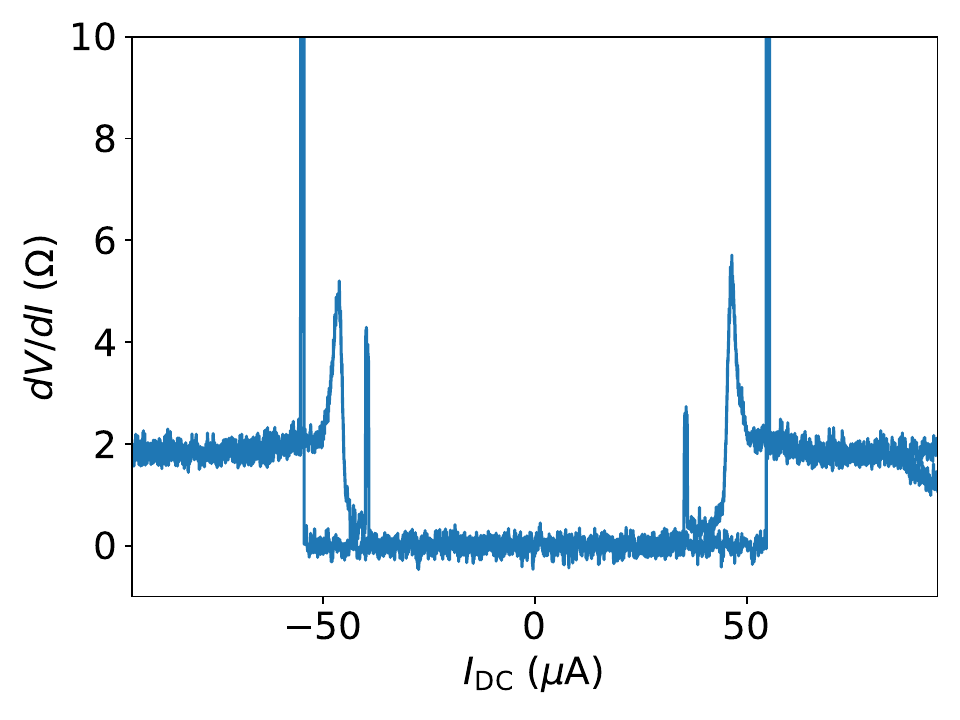}}
	\caption{Data of the second 30$\times$30 four-terminal Josephson junction array. a) Differential resistance-bias current-magnetic field diagram. The oscillations of the critical current are clearly visible. b) Scanning electron microscopy image of the device. c) I-V measurement at 80\;mK. d) Differential resistance calculated from the data in c). \label{fig:SUPP_2nd_Array}}
\end{figure}

In addition to the 30$\times$30 4TJJ array presented in the main text, there was a second device on the same sample. It equals the first in every aspect. Here, we present the data of the second 4TJJ array (see Fig.~\ref{fig:SUPP_2nd_Array}).\\
The critical current oscillates with magnetic field in the same way as for the main text array. With a DC bias of 30\;$\mu$A, that behavior leads to an oscillation of the resistance with magnetic field [see Fig.~\ref{fig:SUPP_Mag_Hyst}~a)]. Similar to the main text device, the frustrated frustration pattern is present. The damping of the oscillations of resistance and critical current can be reproduced.\\ 
At zero magnetic field and a temperature of 80\;mK, the critical current is comparable to the main text device with $I_\mathrm{c} = 59.75\;\mu$A [see Fig.~\ref{fig:SUPP_2nd_Array}~c)~and~d)]. The differential resistance is lower with $R = 1.75\;\Omega$ close to the maximum applied current of 100\;$\mu$A.

\section{Reference two-terminal Josephson junction}\label{sec:SUPP_2TJJ}

\begin{figure}[h]
	\centering
	\includegraphics[width=0.55\textwidth]{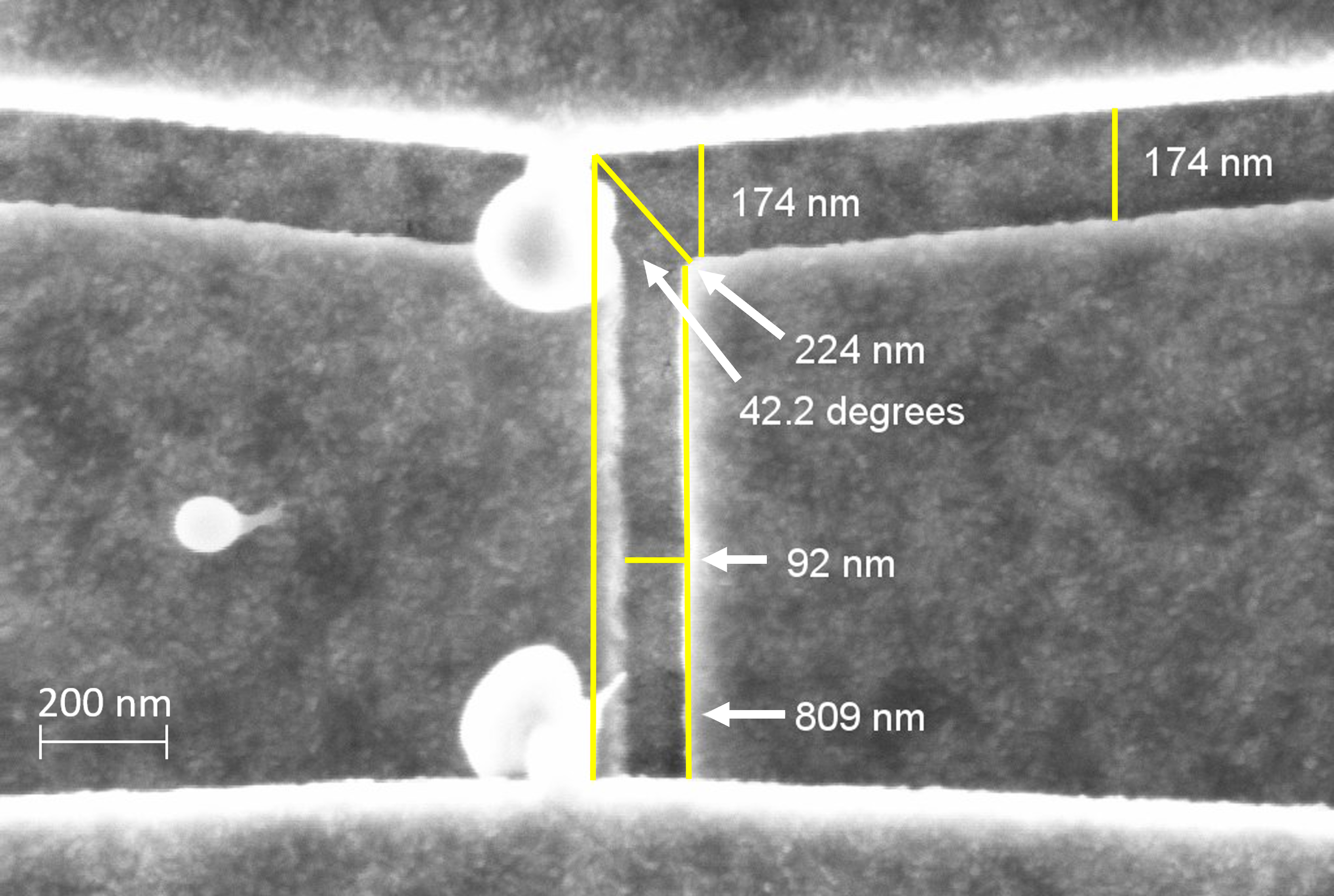}
	\caption{Scanning electron microscope images of the reference two-terminal Josephson junction.\label{fig:SUPP_RefJJ_SEM}}
\end{figure}

On the chip with the presented arrays, a reference two-terminal Josephson junction has been fabricated in the same way as the arrays (see Fig.~\ref{fig:SUPP_RefJJ_SEM}). The weak link length was determined to be 92\;nm and its width is 920\;nm. Although only a width of 809\;nm is indicated in Fig.~\ref{fig:SUPP_RefJJ_SEM}, Pt and Nb material under the edge of the bottom Si$_3$N$_4$ layer needs to be taken into account. The angle between the substrate surface and Nb beam was 57.7$^\circ$. The 300\;nm-thick-SiO$_2$ layer right at the edge of the trench is etched away, just like under the mask itself. This leaves the 100\;nm-thick-Si$_3$N$_4$ layer under-etched at the edge and gives additional substrate space for the niobium shadow to be deposited. The extent of the Nb shadow under the Si$_3$N$_4$ edge can be derived from the top shadow of Fig.~\ref{fig:SUPP_RefJJ_SEM} (174\;nm). The bottom shadow has a smaller distance from its edge than the top shadow because it is delimited by the lower edge of the Si$_3$N$_4$ layer (instead of the upper edge). That means that the bottom shadow will be $100\;\text{nm} / \text{tan}(57.7^\circ) = 63\;\text{nm}$ shorter than the top shadow. This results in a total bottom shadow of 111\;nm, which has already been added to the width specification of this Josephson junction above (920\;nm).\\
In Fig.~\ref{fig:SUPP_RefJJ_SEM}, the junction is shifted horizontally with respect to the stencil mask [by $174\;\text{nm}\cdot\text{tan}(42.2^\circ) = 158\;\text{nm}$] because latter was rotated against the Nb beam during deposition.\\

\begin{figure*}[h]
	\centering
	\subfigure[]{\includegraphics[width=0.49\textwidth]{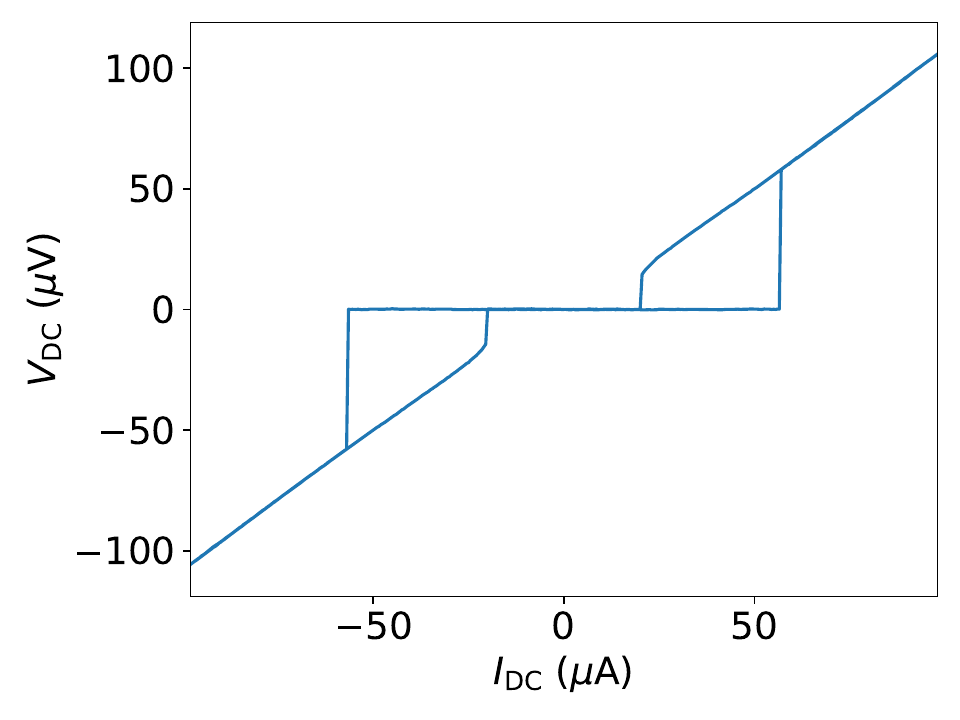}}	\subfigure[]{\includegraphics[width=0.49\textwidth]{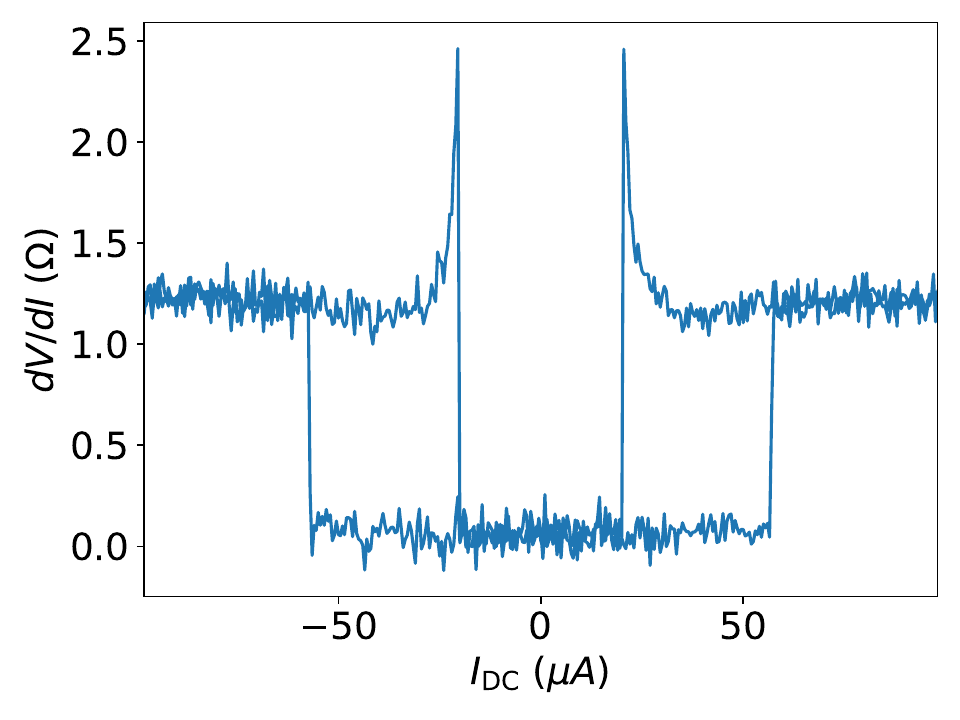}}
	\caption{Measurement data of the reference two-terminal Josephson junction. a) DC IV curve with a critical current is 56.51\;$\mu$A. b) Differential resistance measured with the lock-in measurement technique. It is 1.21$\;\Omega$, close to the maximum applied current of 100\;$\mu$A.\label{fig:SUPP_RefJJ_data}}
\end{figure*}

Figure \ref{fig:SUPP_RefJJ_data} shows the voltage and differential resistance dependent on the current. The device has a critical current of $I_\text{c} = 56.51\;\mu$A and a differential resistance of $R = 1.21\;\Omega$. The Josephson energy of the junction is $E_\mathrm{J} = \Phi_0 I_\mathrm{c}/2\pi = 116$\;meV. Using a relative permittivity of one (because platinum is a metal) and the geometrical values given above (length of a SC contact is 600\;$\mu$m), the coplanar junction capacitance is $C = 5.6\cdot10^{-17}\;$F.\\
Using the geometric device values, the device resistivity is $\rho = 242$\;n$\Omega$m and its critical current density is $j_\mathrm{c} = 3.071$\;kA/mm$^2$.\\
The Stewart-McCumber parameter for these values (assuming $R_N \approx R$) is $\beta_\text{c} = 2eI_\text{c}R_\text{N}^2C/\hbar \approx 1.4\cdot10^{-5}$. Therefore, the junction is in the over-damped regime. Although the junction is over-damped, the current-voltage characteristic of the device shows hysteretic behavior [see~Fig.~\ref{fig:SUPP_RefJJ_data}~a)]. This can be explained by an increase in electron temperature in the resistive state \cite{courtois_origin_2008a}.

\section{Magnetic hysteresis of the resistance pattern}
\begin{figure}[h]
	\centering
	\subfigure[Main text 4TJJ array.]{\includegraphics[width=0.49\textwidth]{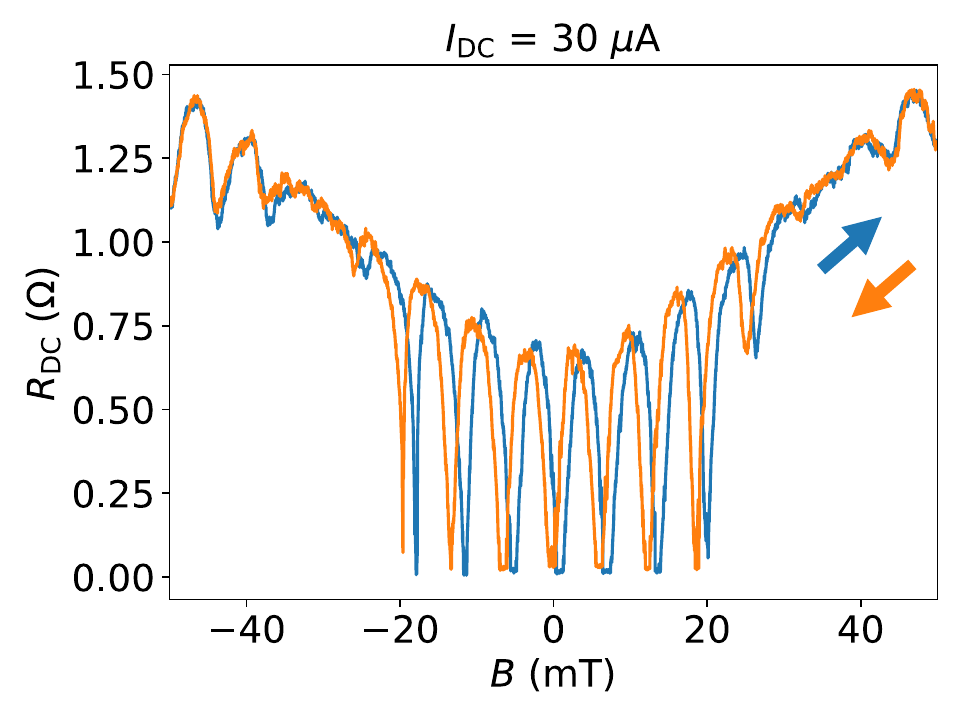}}
	\subfigure[Second 4TJJ array.]{\includegraphics[width=0.49\textwidth]{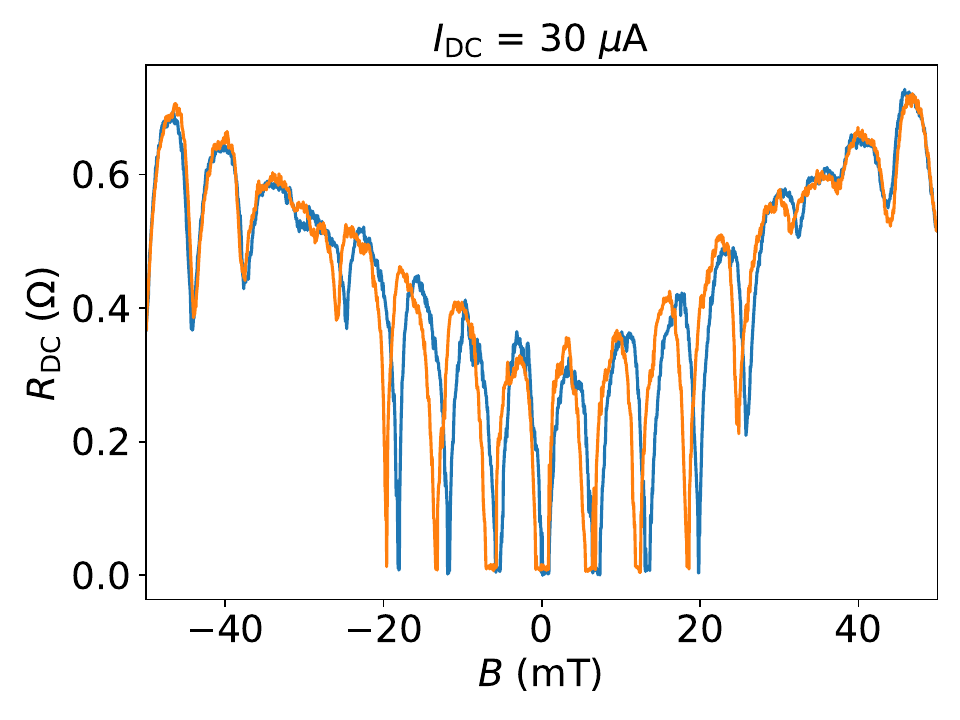}}\\
	\subfigure[Reference 2TJJ.]{\includegraphics[width=0.49\textwidth]{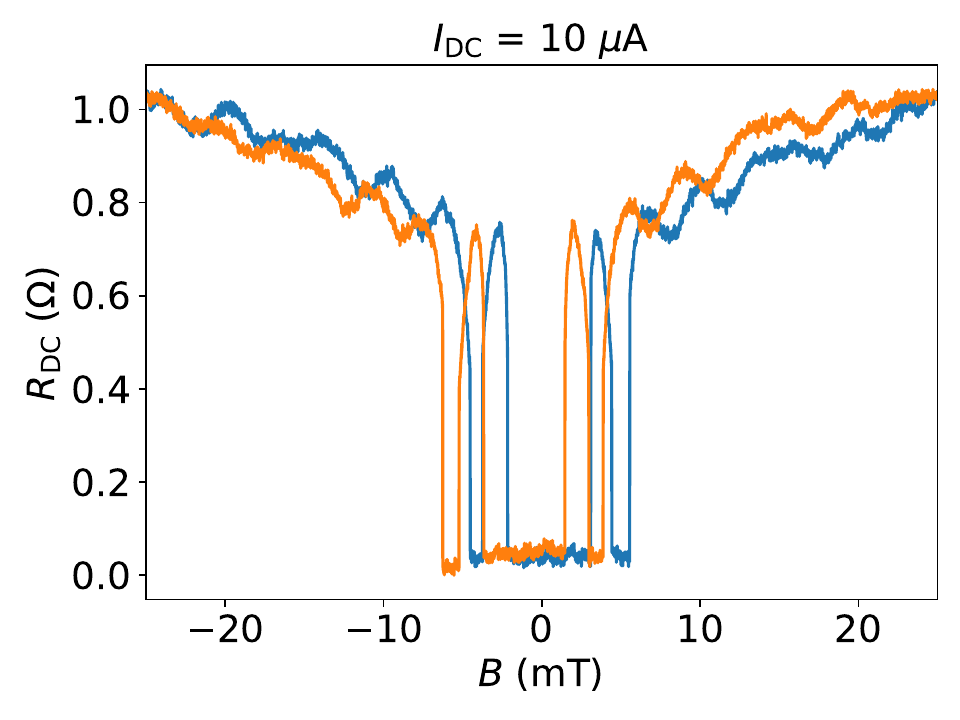}}
	\caption{Resistance dependent on magnetic field for: (a) the main text 4TJJ array, (b) the second 4TJJ array, (c) and the reference two-terminal Josephson junction. The blue lines indicate a magnetic sweep direction from negative to positive and the orange lines from positive to negative. All devices show hysteretic behavior with magnetic field. The two upper resistance oscillations stem from the frustration pattern of the array, the lower resistance oscillation stems from the Fraunhofer effect of the two-terminal Josephson junction. \label{fig:SUPP_Mag_Hyst}}
\end{figure}

A magnetic hyteresis has been observed when measuring the resistance with magnetic field. The resistance oscillation pattern is shifted for different sweep directions of the magnetic field (see Fig.~\ref{fig:SUPP_Mag_Hyst}). The shift has been observed in all devices measured: In the arrays as well as in the two-terminal Josephson junction. Therefore, we rule out that the effect is device-specific. A possible reason is the screening of magnetic field by present Abrikosov vortices.\\
In the main text, we remove the hysteresis by showing only one sweep direction of the magnetic field, namely from negative to positive, and shift the data by 0.9\;mT to the negative side.\\
\\
The period of the Fraunhofer effect for the reference two-terminal Josephson junction (2TJJ) in Fig.~\ref{fig:SUPP_Mag_Hyst}~c) does not fit to the weak link area given in suppl. section~\ref{sec:SUPP_2TJJ}. Possible reasons for this are an unequal current distribution, flux focusing, or the presence of Abrikosov vortices.

\section{Determination of the resistance oscillation period}

\begin{figure}[h]
	\centering
	\includegraphics[width=0.6\textwidth]{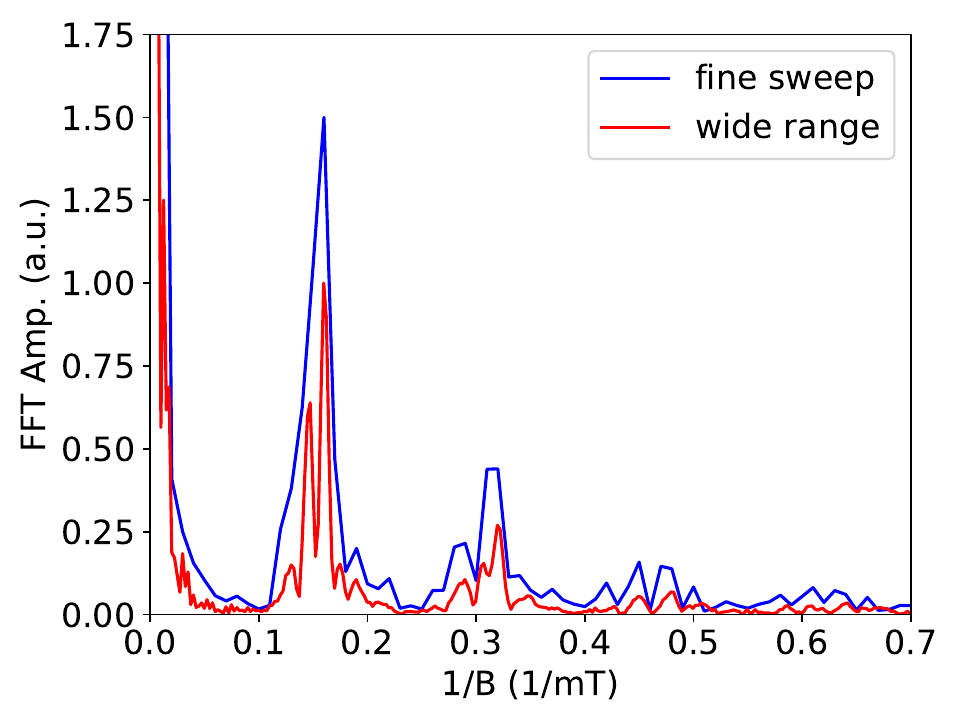}
	\caption{Fast Fourier transforms of the data curves in main text Fig.~\ref{FIG:Array_transport_data}~b) (fine sweep) and c) (wide range). The dominant magnetic field frequency (period) is 0.16\;mT$^{-1}$  (6.25\;mT). The other peaks of the spectrum are higher harmonics of that frequency.\label{fig:SUPP_FFT}}
\end{figure}

For the determination of the oscillation frequencies, fast Fourier transforms have been applied to the data in Fig.~\ref{FIG:Array_transport_data}~b)~and~c) of the main text (see Fig.~\ref{fig:SUPP_FFT}). The frequency spectra of both data curves match. The dominant frequency (period) is 0.16\;mT$^{-1}$ (6.25\;mT). The period of the beating pattern cannot be seen in the spectrum, probably because it is too close to the zero peak.

\section{Description of the array device}\label{sec:SUPP_Array_Device_Description}

\begin{figure*}[h]
	\centering
	\subfigure[]{\includegraphics[width=0.49\textwidth]{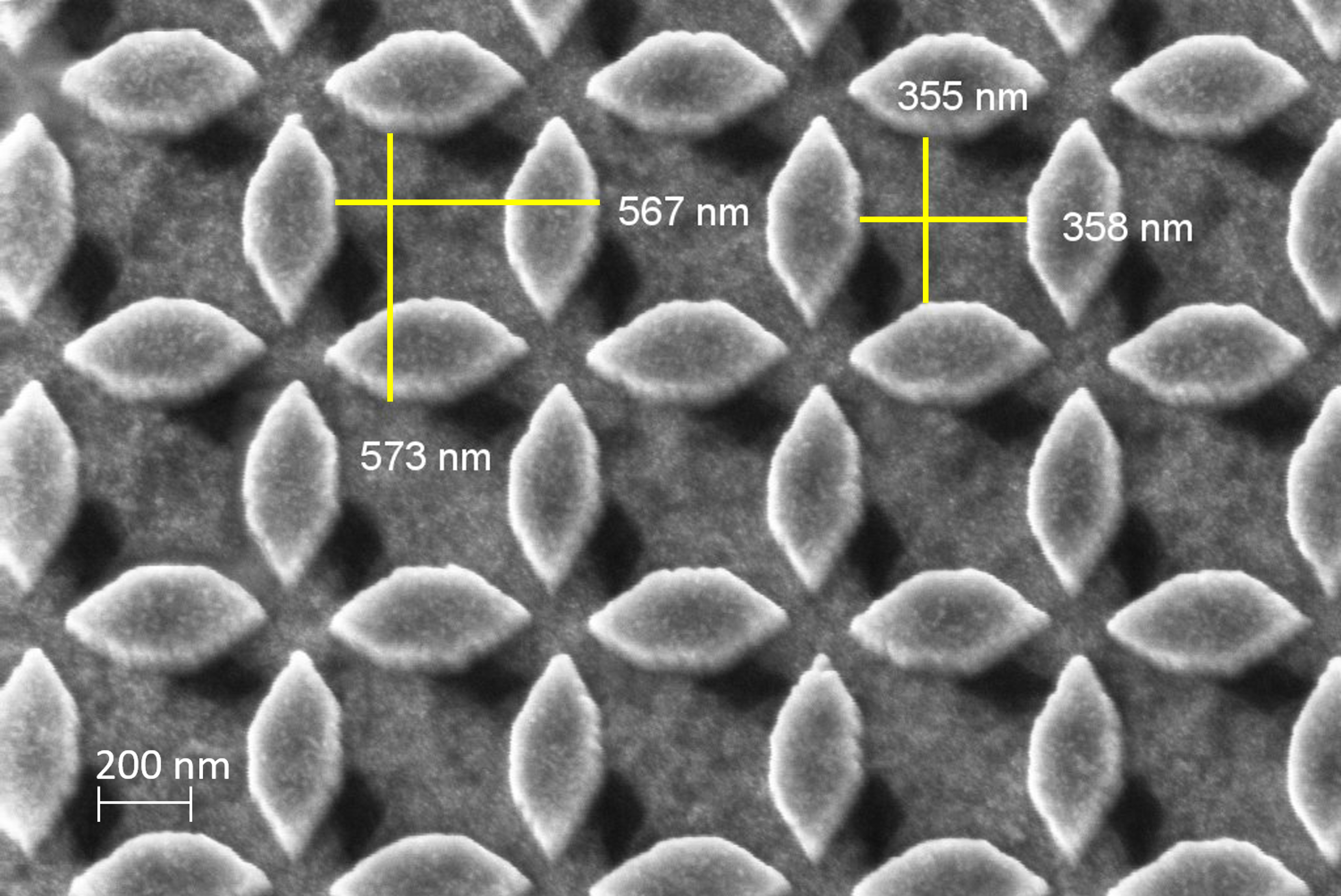}}
	\subfigure[]{\includegraphics[width=0.49\textwidth]{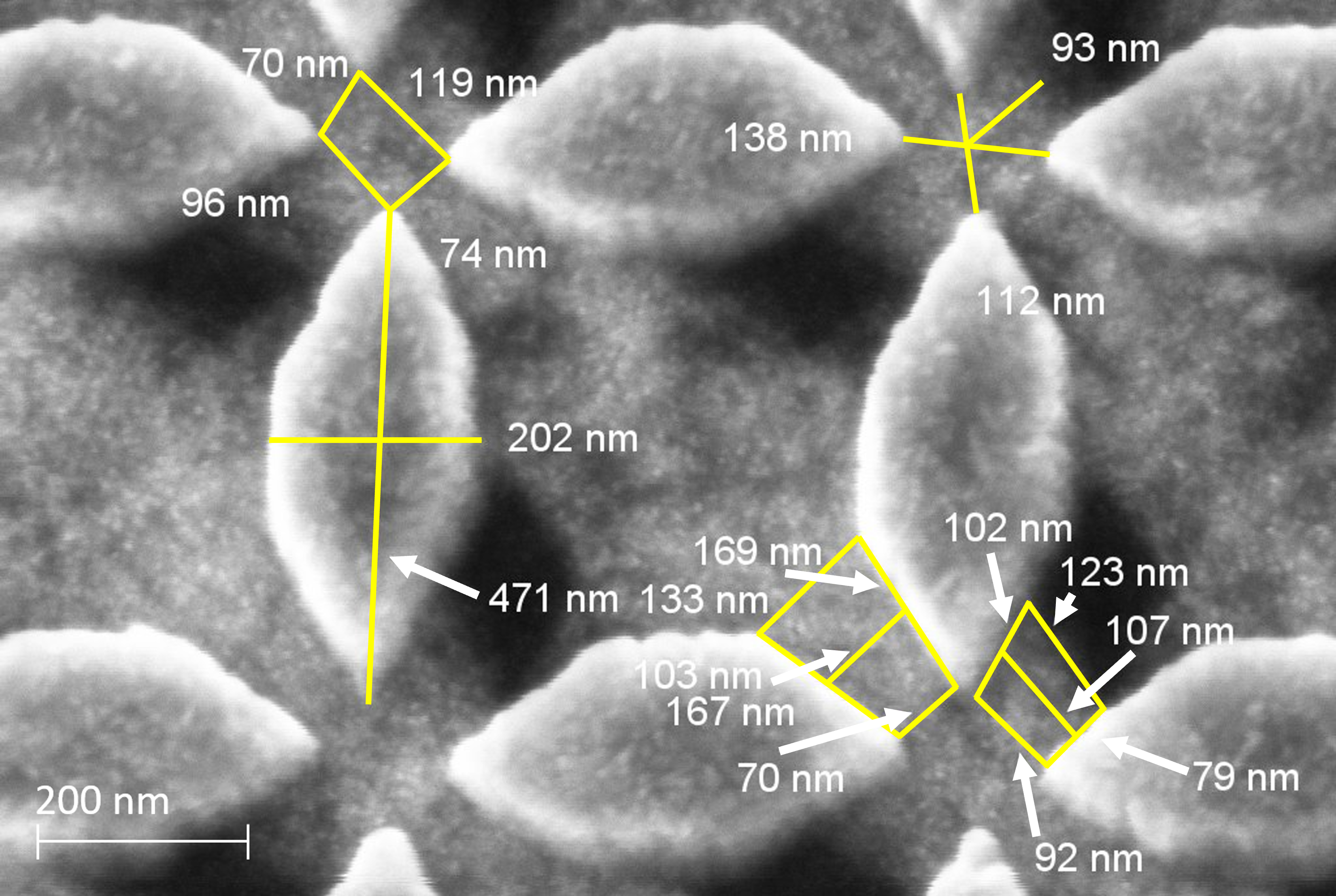}}\\
	\subfigure[]{\includegraphics[width=0.49\textwidth]{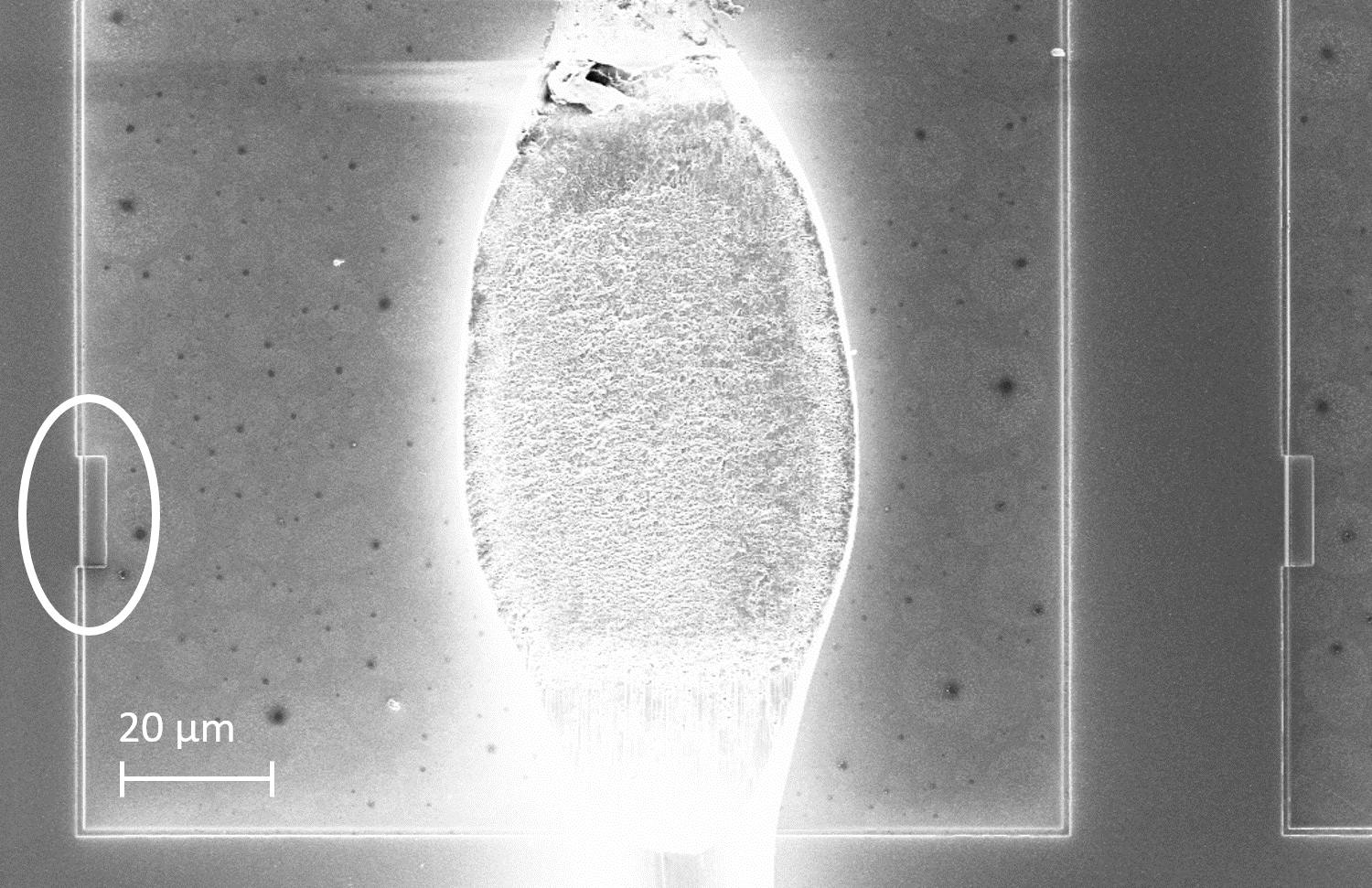}}
	\subfigure[]{\includegraphics[width=0.475\textwidth]{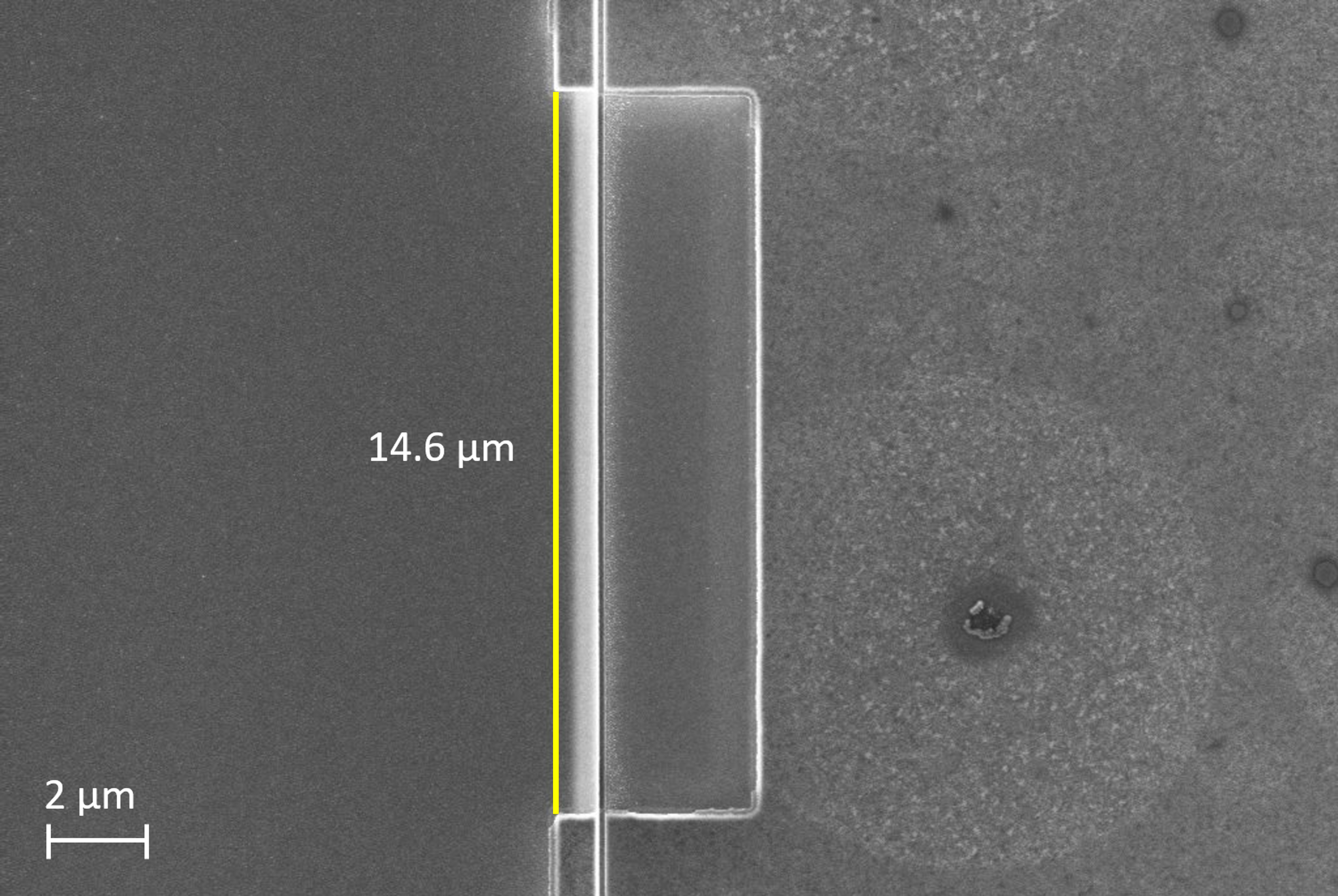}}
	\caption{Scanning electron microscope images of the main text array. a): Description of the lattice geometry. b): Geometry of one 4TJJ in the array. c): Etching of the superconducting edge going around the trench of the device (left). The etchings are located close to the contact endings. A bond wire is visible in the middle. d): Close-up of a corresponding etching. The right side is the contact trench on the Si(111) substrate. The left side is the SiO$_2$ (300\;nm) / Si$_2$N$_3$ (100\;nm) layer around the device.\label{fig:Array_Device_Description}}
\end{figure*}

This section gives a detailed description of the array device in the main text, including geometrical properties of the array lattice (see Fig. \ref{fig:Array_Device_Description}). According to Fig.~\ref{fig:Array_Device_Description} a), the unit cell of the array is a square of size $A_\text{UC} =573\;\text{nm}\cdot567\;\text{nm} = (570\;\text{nm})^2$. The geometry of the weak link area between two neighboring leads of one 4TJJ [see Fig.~\ref{fig:Array_Device_Description}~b)] is described, on average, by the following dimensions: The width of the weak link is 129\;nm and its length is 97\;nm. The superconducting arms in the 4TJJ array have a length of 471\;nm.\\
\\
During fabrication of the device, a trench is etched into the SiO$_2$ (300\;nm) / Si$_2$N$_3$ (100\;nm) layer, down to the Si(111) substrate. On the substrate, the device is fabricated and, afterwards, the whole chip is covered with superconducting material. To minimize disturbance effects, the superconductor around the device is etched away as a last fabrication step, leaving an edge of superconductor of approximately 1\;µm around the trench behind. To prevent this closed loop of superconductor from generating screening currents, the edge is disconnected close to the ends of the device contacts [see Figs.~\ref{fig:Array_Device_Description}~c)~and~d)]. 
\newpage

\section{Characteristic resistively capacitively shunted junction (RCSJ) values between two leads of a 4TJJ in the arrray}

In this section, relevant parameters for the theoretical description of the array are derived from geometrical considerations of the prior section, as well as from measurement data of the reference two-terminal Josephson junction. We assume that critical current density and resistivity, derived in suppl. section \ref{sec:SUPP_2TJJ}, are geometry independent. We then use these values, together with the 4TJJ geometry, to derive capacitance, resistance, and critical current between two leads of one 4TJJ.\\
The coplanar capacitance is calculated using the geometrical values given in suppl. section~\ref{sec:SUPP_Array_Device_Description}. The relative permeability $\epsilon_r$ is assumed to be one, since the weak link is a metal. This results in a capacitance of $C = 2.729\cdot 10^{-18}$\;F. The critical current between two 4TJJ leads is $I_\mathrm{C} = 7.9$\;µA. This results in a Josephson energy of $E_\mathrm{J} = \Phi_0 I_\mathrm{c}/2\pi = 16.2$\;meV. The resistance is equal to $R = 9.1\;\Omega$. The Stewart-McCumber parameter for these values (assuming $R_N \approx R$) is $\beta_\text{c} = 2eI_\text{c}R_\text{N}^2C/\hbar \approx 5.4\cdot10^{-6}$. Therefore, the junction is in the over-damped regime.\\
Although the junction is over-damped, the current-voltage characteristic of the array shows hysteretic behavior [see~Fig.~\ref{fig:SUPP_MISC_IV_base}~a)]. This can be explained by an increase in electron temperature in the resistive state \cite{courtois_origin_2008a}.

\section{Additional measurements of the main text 30$\times$30 4TJJ array}\label{sec:SUPP_Misc_Measurements}
\begin{figure}[h]
	\centering
	\subfigure[]{\includegraphics[width=0.49\textwidth]{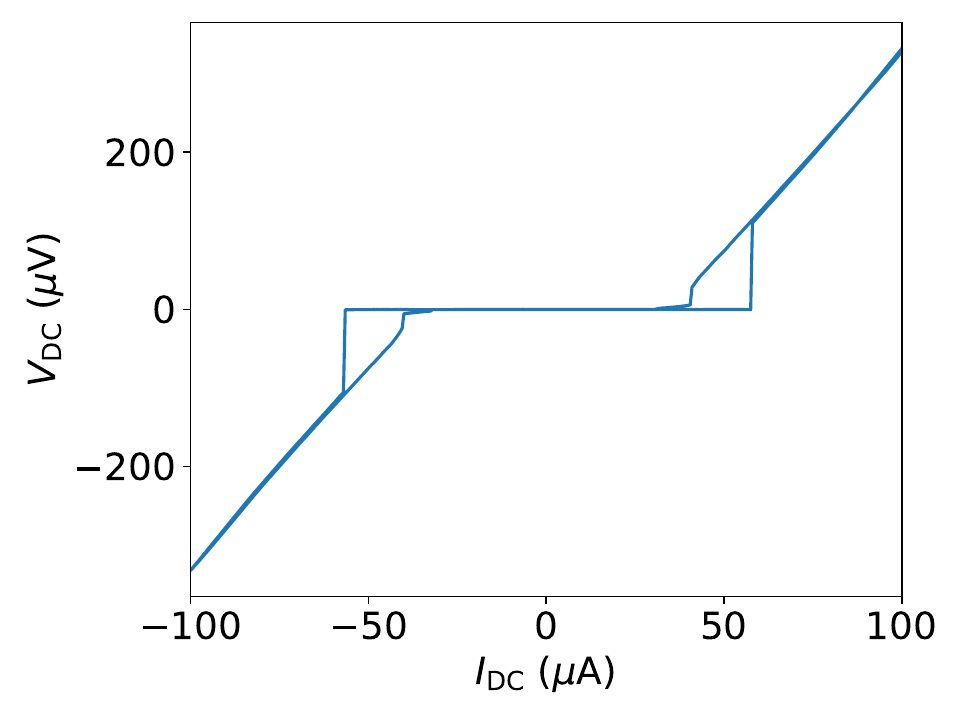}}
	\subfigure[]{\includegraphics[width=0.49\textwidth]{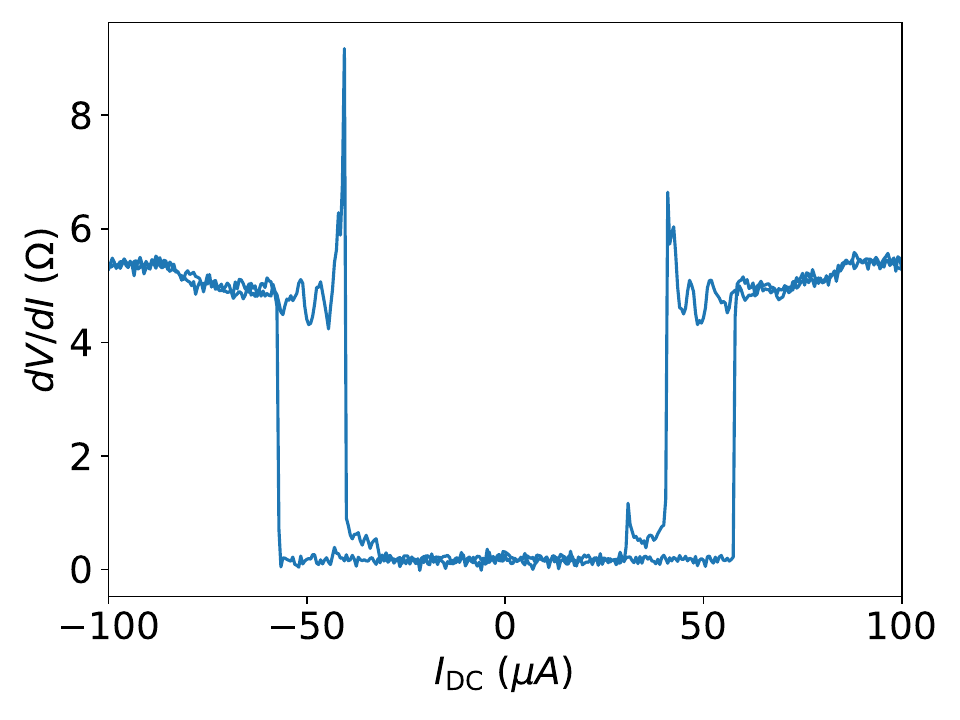}}
	\caption{IV measurements of the array discussed in the main text at 80\;mK. a) The DC IV curve with a critical current of $I_\mathrm{c} = 57\;\mu$A. b) AC differential resistance measured with the lock-in technique. The AC bias was 200\;nA.\label{fig:SUPP_MISC_IV_base}}
\end{figure}

The current-voltage characteristics of the device presented in the main text can be seen in Fig.~\ref{fig:SUPP_MISC_IV_base}. The device has a critical current of $57\;\mu$A and a differential resistance of around 5.5\;$\Omega$ close to the maximum applied current of 100\;$\mu$A.

\begin{figure}[h]
	\centering
	\subfigure[]{\includegraphics[width=0.49\textwidth]{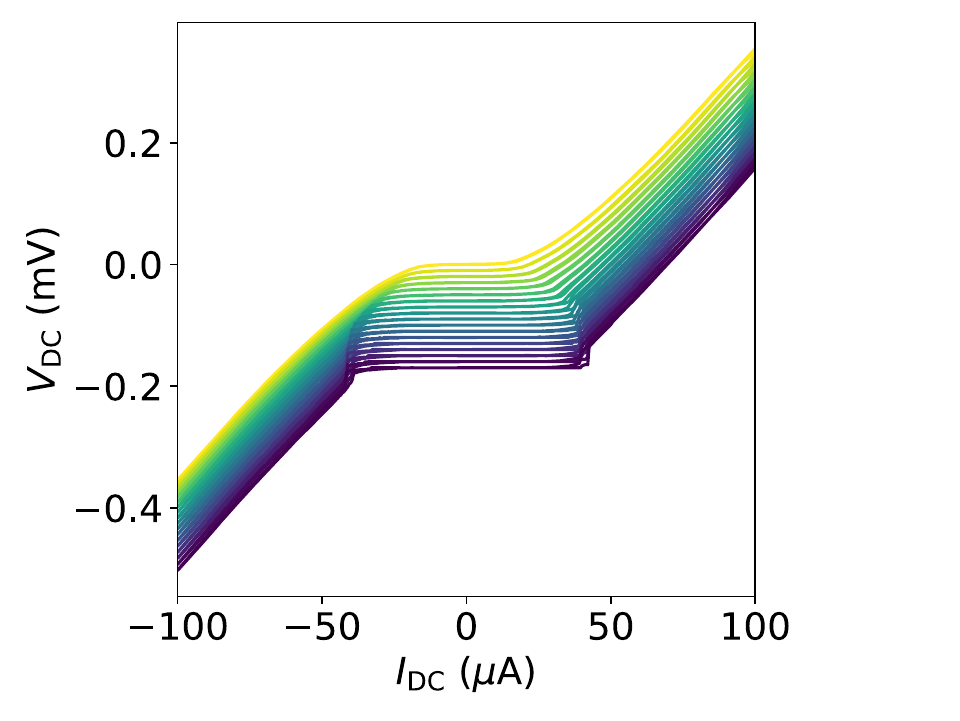}}
	\subfigure[]{\includegraphics[width=0.49\textwidth]{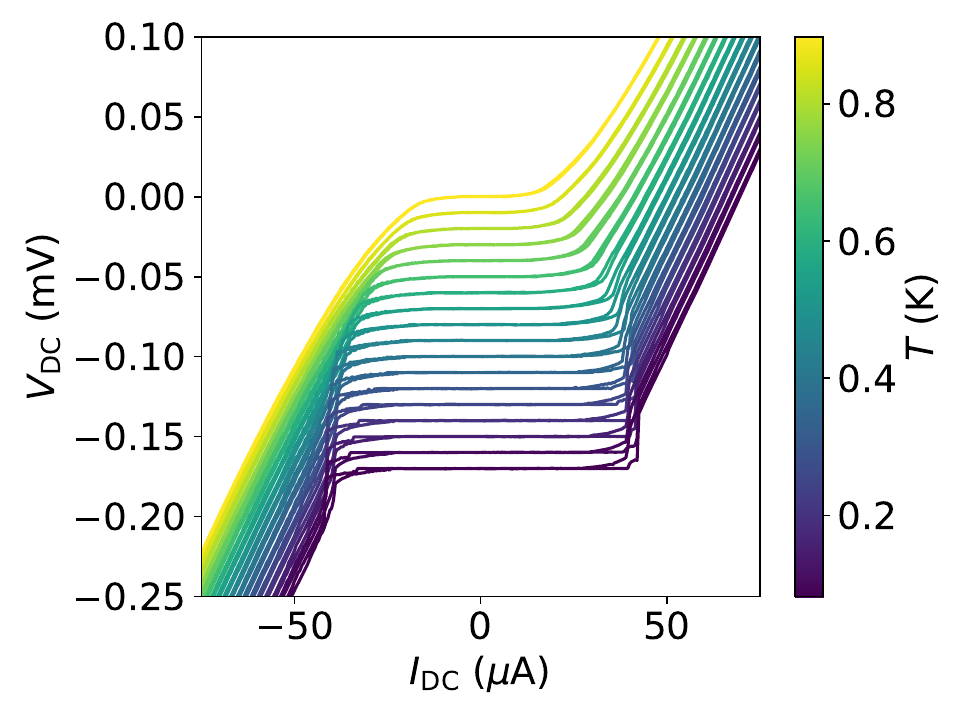}}
	\caption{IV measurements dependent from 70\;mK to 900\;mK. Each measurement is shifted by -0.01\;$\mu$V from the one before. b): Zoom in of a).\label{fig:SUPP_MISC_IV_T}}
\end{figure}

Additional measurements have been performed on the main text 30$\times$30 array device. Figure \ref{fig:SUPP_MISC_IV_T} shows temperature dependent IV curves. The declining critical current with increasing temperature is clearly visible. 
\\

\begin{figure}[h]
	\centering
	\subfigure[$I_\mathrm{AC} = 30\;\mu$A.]{\includegraphics[width=0.49\textwidth]{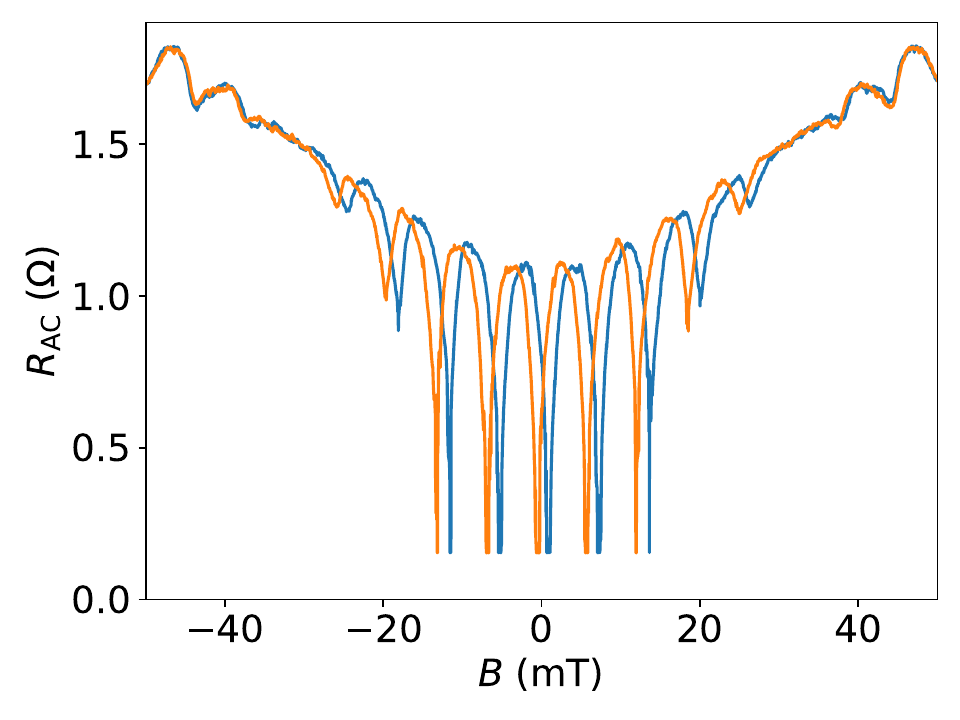}}
	\subfigure[$I_\mathrm{AC} = 40\;\mu$A.]{\includegraphics[width=0.49\textwidth]{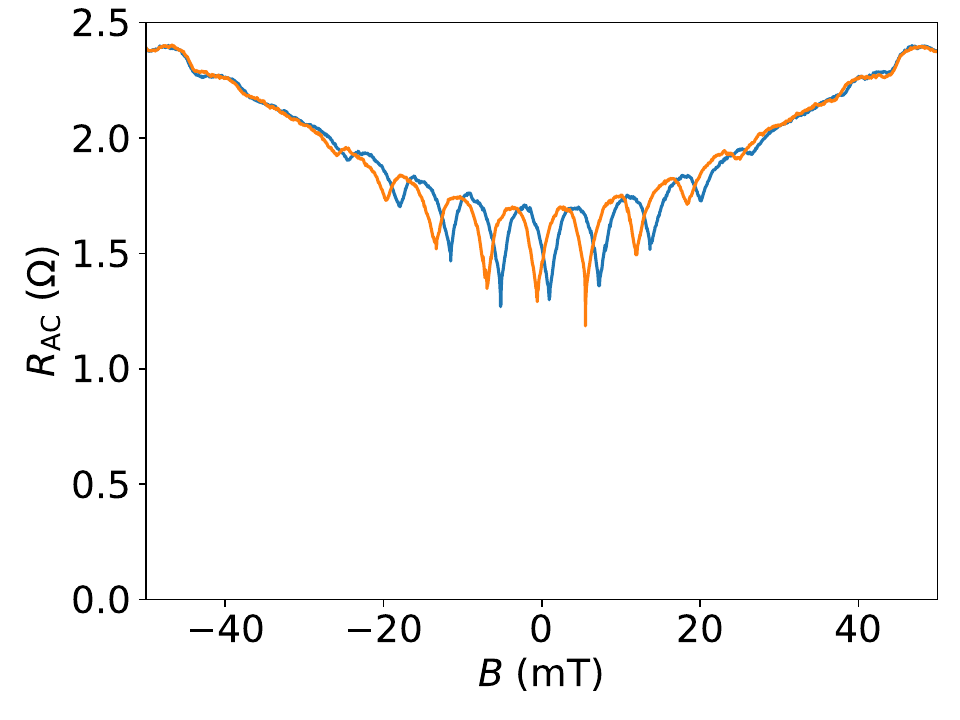}}
	\caption{AC resistance measurement dependent on magnetic field for 30\;$\mu$A (a) and 40\;$\mu$A (b). The blue lines indicate a magnetic sweep direction from negative to positive and the orange lines from positive to negative.\label{fig:SUPP_MISC_R(B)_AC}}
\end{figure}

Figure~\ref{fig:SUPP_MISC_R(B)_AC} shows the same resistance oscillations with magnetic field as before but, here, they were measured with the AC lock-in technique and no DC bias. The resulting resistance is lower than with the DC measurement because the time-dependent sinusoidal signal includes the superconducting  state of the device as well. When comparing Fig.~\ref{fig:SUPP_MISC_R(B)_AC}~a) with Fig.~\ref{fig:SUPP_Mag_Hyst}~a), one can see that the oscillations are identical in magnetic field, although the peak pattern is more symmetrical with the AC method.\\
\\

\begin{figure}[h]
	\centering
	\subfigure[]{\includegraphics[width=0.49\textwidth]{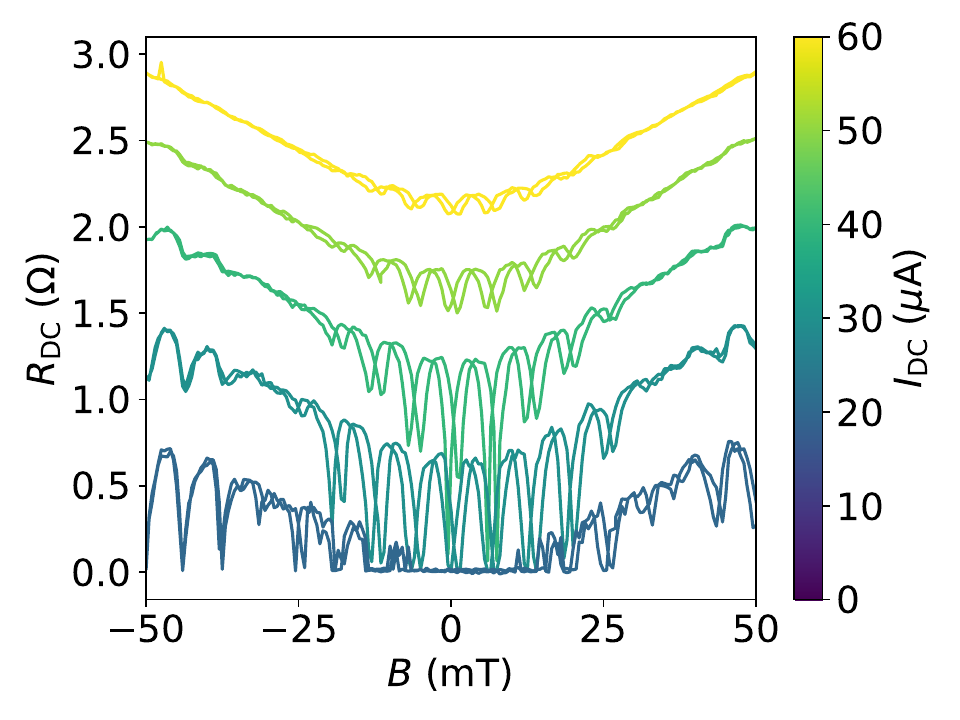}}
	\subfigure[]{\includegraphics[width=0.49\textwidth]{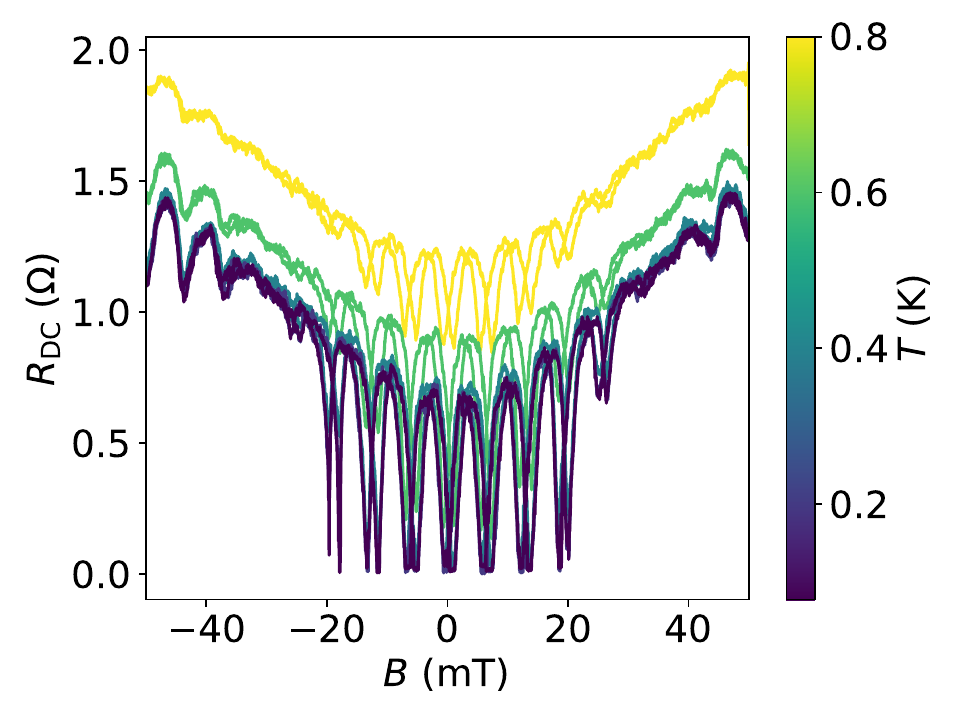}}
	\caption{DC resistance measurements dependent on magnetic field. a): At currents between 20\;$\mu$A and 60\;$\mu$A. b): At a DC bias of $I_\mathrm{DC} = 30\;\mu$A and temperatures between 70\;mK and 800\;mK.\label{fig:SUPP_MISC_R(B)_DC_I_T}}
\end{figure}

Figure \ref{fig:SUPP_MISC_R(B)_DC_I_T} a) shows resistance measurements dependent on magnetic field and DC current. At higher currents, superconductivity is not realized at any magnetic field and the oscillation amplitudes decline.\\
Figure \ref{fig:SUPP_MISC_R(B)_DC_I_T} b) shows resistance measurements dependent on magnetic field and temperature. At higher temperatures, superconductivity is not realized at any magnetic field and the oscillation amplitudes decline.

\section{Details of the resistively capacitively shunted junction model calculations}

We here summarize the model calculations for the theoretical discussion of the main text.
We describe the multiterminal Josephson junction array in the overdamped regime (discarding capacitive contributions), we get at each node $j=(j_x,j_y)$ (where $j_{x,y}=1,\ldots,J_{x,y}$) the equation of motion,
\begin{equation}
	\frac{1}{2eR}\sum_{\langle j^\prime\rangle_j}\left(\dot{\phi}_j-\dot{\phi}_{j^\prime}\right)=-I_c\sum_{\langle j^\prime\rangle_j}\sin(\phi_j-\phi_{j^\prime}+a_{j,j^\prime})+\sum_{\langle j^\prime\rangle_j}\eta_{j,j^\prime}(t)\ ,
\end{equation}
with parallel shunt resistance $R$ and critical current $I_c$. The vector potential $a_{j,j^\prime}$ accounts for the magnetic flux piercing the plaquettes of the array (which will be defined in detail futher below). While we assume zero temperature in the device for the discussion in the main text, we here include for completeness the current noise term $\eta_{j,j^\prime}$ due to thermal fluctuations from the resistor connecting nodes $j$ and $j^\prime$, where 
\begin{equation}\label{eq_eom_array}
	\langle \eta_{i,i^\prime}(t)\eta_{j,j^\prime}(t^\prime)\rangle=\frac{2k_BT}{R}\delta_{i,j}\delta_{i^\prime, j^\prime}\delta(t-t^\prime)\ .
\end{equation}
The notation $\sum_{\langle j^\prime \rangle_j}$ indicates that the sum $j^\prime$ goes over nearest neighbours of $j$. That is, for $j=(j_x,j_y)$, its neighbours are $(j_x\pm 1,j_y)$ and $(j_x,j_y\pm 1)$.

The system further has boundary conditions. At the top and bottom boundaries ($j_y=1$ and $j_y=J_y$), the nodes simply terminate, such that the nearest neighbour sum only has three instead of the usual four terms. On the left hand side, $j_x=1$, the array connects to ground. This connection can be taken into account by artifically extending the lower bound of the $j_x$ index to $0$ (instead of $1$) and setting $\phi_{0,j_y}=0$. The right-hand side on the other hand ($j_x=J_x$) connects to the current-biased terminal. Here, we artificially extend of the upper bound of $j_x$ from $J_x$ to $J_x+1$ and replace in the sums $\phi_{(J_x+1,j_y)}=\phi_b$, where $\phi_b$ is the superconducting phase of the current-biased terminal. This phase satisfies an equation of motion of its own,
\begin{equation}
	\frac{1}{2eR}\sum_{j_y}\left(\dot{\phi}_b-\dot{\phi}_{(J_x,j_y)}\right)=-I_c\sum_{j_y}\sin\left(\phi_b-\phi_{(J_x,j_y)}+a_{b,(J_x,j_y)}\right)+\sum_{j_y}\eta_{b,(J_x,j_y)}+I \ ,
\end{equation}
where the current bias $I$ appears as an additional force term.
The voltage built up across the device is defined through the phase velocity $\dot{\phi}_b$. The dc resistance across the total device is thus given as
\begin{equation}
	R_\text{dc}=\lim_{\tau\rightarrow \infty}\frac{\phi_b(\tau)-\phi_b(0)}{2e\tau I}\ .
\end{equation}
For sufficiently long measurement times $\tau$, this quantity no longer depends on the initial conditions at $t=0$.

The vector potentials must be chosen such that sums along counter-clockwise loops yield the flux through a plaquette $(j_x,j_y)$ (note that plaquette indices are of course shifted with respect to node indices). Defining that flux as $f_{(j_x,j_y)}$, then the vector potential needs to satisfy
\begin{equation}
	f_{(j_x,j_y)}=a_{(j_x,j_y),(j_x+1,j_y)}+a_{(j_x+1,j_y),(j_x+1,j_y+1)}+a_{(j_x+1,j_y+1),(j_x,j_y+1)}+a_{(j_x,j_y+1),(j_x,j_y)}\ .
\end{equation}
For the checker board model, the flux yields either $f$ or $f^\prime$ if $j_x+j_y$ is even or odd, respectively.
We further define $a_{j_x,j_y}^{(y)}=a_{(j_x,j_y),(j_x,j_y+1)}$ and $a_{j_x,j_y}^{(x)}=a_{(j_x,j_y),(j_x+1,j_y)}$, and choose $a_{j_x,j_y}^{(y)}=0$. Using $a_{jj^\prime}=-a_{j^\prime j}$, we find a valid gauge as
\begin{equation}
	a_{(j_x,j_y)}^{(x)}=\sum_{j_x^\prime=1}^{j_x}f_{(j_x^\prime,j_y)}\ ,
\end{equation}
which corresponds to a lattice version of the Landau gauge.

In the absence of the noise term, the system can be solved as a first order differential equation, which for all numerical results shown in the main text is done directly with the NDSolve routine on Mathematica.

In Eq.~\eqref{eq_Rdc_null} in the main text, we provide an exact expression for the special case of $f,f^\prime\in\mathbb{Z}$ (equivalent to $f=f^\prime =0$). We here derive this expression. For the phases within the array, we choose the ansatz
\begin{equation}
	\phi_{(j_x,j_y)}=\frac{j_{x}}{J_{x}+1}\phi_{b}\ .
\end{equation}
At zero temperature and $f=f^\prime=0$, this ansatz immediately solves Eq.~\eqref{eq_eom_array} at all nodes. The remaining equation of motion is simply that of $\phi_b$, which is simply given as
\begin{equation}
	\frac{1}{2eR}\frac{1}{J_{x}+1}\dot{\phi}_{b}=-I_{c}\sin\left(\frac{1}{J_{x}+1}\phi_{b}\right)+\frac{I}{J_{y}}\ .
\end{equation}
This is simply the equation of motion for a single Josephson junction with voltage and bias current rescaled by the lattice size parameters $J_{x}+1,J_y$. The resulting dc resistance of this reduced equation is well known. Carefully accounting for the rescaling, we arrive at Eq.~\eqref{eq_Rdc_null}.

\section{Comments on stability of frustrated frustration patterns in vortex language}

In the main text, we pointed out that resistive dips at integer $f+f^\prime$ can be stable if $f$ and $f^\prime$ are individually close (but not equal) to integer, even though the regular frustration pattern is absent. We argued that the increased stability of those features can be understood in terms of loop currents. Here we provide an alternative picture based on the more commonly used language of vortices and the Berezinskii–Kosterlitz–Thouless (BKT) transition. The energy of the classical 2D Josephson junction array in terms of the number of vortices $v$ per plaquette is given by (see, e.g., Ref.~\cite{penner_resistivity_2023} for details),
\begin{equation}
	H=\sum_{i,j}(v_{i}-f_i)U_{ij}(v_j-f_j)\ ,
\end{equation}
where the indices $i,j$ address the plaquette positions in 2 dimensions (i.e., similar to the previous section, both $i$ and $j$ implicitly encode a pair of $x$ and $y$ coordinates), and $U_{ij}$ represents the typical logarithmic long-range interaction. For regular square lattices $f_i=f$ (i.e., all plaquettes experience the same flux), the well-known BKT picture arises. At zero flux $f=0$, the zero vortex state $v_j=0$ (which is insulating in that no vortices are available to move) corresponds to the ground state, whereas states with finite vortex numbers or vortex pairs (conducting states) are at low temperatures inaccessible due to a finite energy gap, leading to the BKT transition. At finite flux, e.g., $f>0$, the system reacts in such a way that the new ground state can be to a good approximation described by a state with a finite vortex density of $\sim f$ vortices per plaquette, in order to screen the applied external flux, where (as is well-known) the propensity of the ground state to have ordered (and thus likewise insulating) vortex configurations or many freely moving vortex configurations depends on whether $f$ is a rational number or not.

For the  checker board flux texture on the other hand, the situation is different in the following crucial aspect. The special points of integer $f+f^\prime$ can be alternatively regarded as cancelling neighbouring fluxes $f^\prime=-f$ (up to modulo one). At these special points we still have the zero vortex state $v_j=0$ as the gapped ground state because averaged over the whole system, no flux is applied, and hence no finite vortex density occurs due to screening. We therefore arrive again at the superconducting BKT phase, notably, at generally irrational values of $f$. We emphasise that these zero vortex states are by no means trivial. Consider the special (rational) case of $f=1/2$ and $f^\prime=-1/2$, with the zero vortex ground state. With a simple unitary transformation (increasing the vortex number on every $-1/2$ plaquette by one), we map to the regular square lattice $f=f^\prime=1/2$ with the well-known ground state given by an alternating $0,1$ vortex configuration. Consequently, $f=-f^\prime=1/2$ with the zero vortex ground state and $f=f^\prime=1/2$ with alternating $0,1$ vortex ground state are equivalent. Now, the difference to the case where $f=-f^\prime$ but $f$ irrational (or in fact any rational or irrational number other than $f=0$ or $f=1/2$) is that the correspondence to the regular lattice $f=f^\prime$ is no longer exact (as there exists no unitary mapping due to $v_i \in \mathbb{Z}$). For values close to $1/2$, there still exists a unitary map to a system approximately equivalent to $f=f^\prime =1/2$, this correspondence grows weaker and weaker as we move away from $1/2$, which at the value of $1/4$ or lower leads to a cross over, where the checker board system with $f=-f^\prime$ more closely imitates the regular $f=f^\prime =0$ system instead of $f=f^\prime =1/2$. Regarding stability, this means that the points of $f+f^\prime$ which are closer to $f,f^\prime$ integer (than half-integer) have a higher stability, and they can be observed at higher current bias (or higher temperature) before the half integer resistance dips kick in. This qualitative picture is in agreement with the findings of the resistively capacitively shunted junction model calculation, see main text, and previous supplementary section.

\section{Estimation of $\beta$ and the central weak link area}
\begin{figure}[h!]
	\centering
	\includegraphics[width=0.45\textwidth]{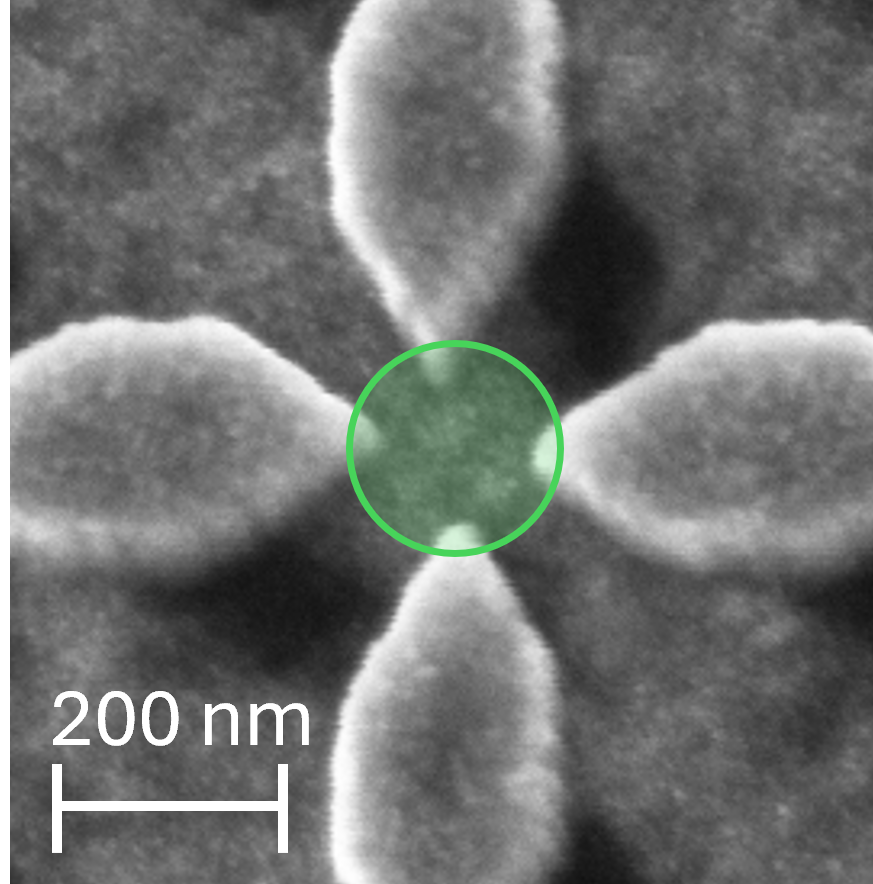}
	\caption{A 4TJJ of the main text array device presented in the main text with the estimated central weak link area sketched in green. Shape and position of the weak link area were chosen freely, its area equals the one estimated by the beating pattern, $(165\;\text{nm})^2$.}
	\label{fig:SUPP_weak_link_area}
\end{figure}

According to the theoretical discussion in the main text, the parameter $\beta$, the dip periodicity ($p_\text{dips}$), and the beating periodicity ($p_\text{beating}$), in frustration space, are given by:
\begin{equation}
	\beta = \frac{f}{f'} = \frac{A}{A'},
\end{equation}
\begin{equation}
	p_\text{dips}  = \frac{\beta}{\beta + 1},
\end{equation}
\begin{equation}
	p_\text{beating} \approx \beta.
\end{equation}
One can use the the difference in periodicities $\Delta p = p_\text{beating} - p_\text{dips}$ to estimate $\beta$:
\begin{equation}
	\Delta p \approx \beta - \frac{\beta}{\beta + 1}.
\end{equation}
For our estimate $\Delta p$ was determined by counting the number of resistance dips (10 dips) between the two beating points in Fig.~\ref{FIG:Array_transport_data} c). This yields $\beta \approx 10.9$.\\
The central weak link area of the 4TJJs is connected to the unit cell area of the array ($A_\text{uc}$) via $A' = A/\beta = (A_\text{uc} - A')/\beta \Rightarrow$ 
\begin{equation}
	A' = \frac{A_\text{uc}}{(\beta+1)}.
\end{equation}
Thus, the 4TJJs in the array have an estimated weak link area of $A' \approx (165\;\text{nm})^2$.\\
For comparison, such a central weak link area is sketched on one 4TJJ in Fig. \ref{fig:SUPP_weak_link_area}, assuming the weak link region to be a perfect circle with the corresponding area of $(165\;\text{nm})^2$. As can be seen, a shape of this size is compatible with the 4TJJ geometry.

\let\oldaddcontentsline\addcontentsline
\renewcommand{\addcontentsline}[3]{}

\let\addcontentsline\oldaddcontentsline

\end{document}